\documentclass[a4paper,11pt]{article}
\pdfoutput=1 

\usepackage{jheppub} 
\usepackage{mathrsfs,graphicx,rotating,amsmath,amsfonts,mathtools,booktabs,amssymb,wasysym,caption}
\usepackage{hyperref}
\usepackage{slashed}
\usepackage{natbib}
\usepackage[table,xcdraw,dvipsnames]{xcolor}
\usepackage{graphicx}
\usepackage{bbold}
\usepackage[utf8x]{inputenc}
\usepackage[english]{babel}
\usepackage{multirow,multicol}
\usepackage{epstopdf}
\usepackage{bbm}
\usepackage{appendix}
\usepackage{libertine}
\usepackage{braket}
\usepackage{wasysym}
\usepackage{empheq}
\usepackage{cancel}
\usepackage{enumitem}
\usepackage{mathrsfs}
\usepackage{lmodern}
\usepackage{tabularx}
\usepackage{multicol}
\usepackage{color}
\usepackage{mathtools}

\newcommand{\Q}{Q}

\renewcommand\[{\left[}

\newcommand{\be}{\begin{equation}}
\newcommand{\ee}{\end{equation}}

\title{\boldmath The connection between nonzero density and spontaneous symmetry breaking for interacting scalars}

\author[a]{Alberto Nicolis,}
\author[a]{Alessandro Podo,}
\author[b,c]{Luca Santoni}

\affiliation[a]{Department of Physics, Center for Theoretical Physics, \\Columbia University, New York, 538 West 120th Street, NY 10027, USA}
\affiliation[b]{ICTP, International Centre for Theoretical Physics,  \\ Strada Costiera 11, 34151, Trieste, Italy}
\affiliation[c]{Universit\'e Paris Cit\'e, CNRS, Astroparticule et Cosmologie, 10 Rue Alice Domon et L\'eonie Duquet, F-75013 Paris, France}

\emailAdd{a.nicolis@columbia.edu}
\emailAdd{ap3964@columbia.edu}
\emailAdd{santoni@apc.in2p3.fr}

\abstract{We consider ${\rm U}(1)$-symmetric scalar quantum field theories at zero temperature.
At nonzero charge densities, the ground state of these systems is usually assumed to be a superfluid phase, in which the global symmetry is spontaneously broken along with Lorentz boosts and time translations. We show that,  in $d>2$ spacetime dimensions, this expectation is always realized at one loop for arbitrary non-derivative interactions, confirming that the physically distinct phenomena of nonzero charge density and spontaneous symmetry breaking occur simultaneously in these systems. We quantify this result by deriving universal scaling relations for the symmetry breaking scale as a function of the charge density, at low and high density.
Moreover, we show that the critical value of $\mu$ above which a nonzero density develops coincides with the pole mass in the unbroken, Poincar\'e invariant vacuum of the theory. The same conclusions hold non-perturbatively for an ${\rm O}(N)$ theory with quartic interactions in $d=3$ and $4$, at leading order in the $1/N$ expansion. We derive these results by computing analytically the zero-temperature, finite-$\mu$ one-loop effective potential, paying special attention to subtle points related to the $i\varepsilon$ terms. We check our results against the one-loop low-energy effective action for the superfluid phonons in  $\lambda \phi^4$ theory in $d=4$ previously derived by Joyce and ourselves, which we further generalize to arbitrary potential interactions and arbitrary dimensions. As a byproduct, we find analytically the one-loop scaling dimension of the lightest charge-$n$ operator for the $\lambda \phi^6$ conformal superfluid in $d=3$, at leading order in $1/n$, reproducing a numerical result of Badel et al. For a $\lambda \phi^4$ superfluid in $d=4$, we also reproduce the Lee--Huang--Yang relation and compute relativistic corrections to it. Finally, we discuss possible extensions of our results beyond perturbation theory.}

\begin{document}
\maketitle
\flushbottom

\section{Introduction}

When it comes to conserved charges and the associated symmetries in Quantum Field Theory (QFT), there is a somewhat implicit expectation that having a zero-temperature state with nonzero density for a given charge goes hand in hand with the spontaneous breaking of the associated symmetry. However, these two properties are conceptually different \cite{Nicolis:2011pv}, and in fact there exist physical systems for each possible combination. For example:
\begin{enumerate}
\item
Zero charge density and no spontaneous symmetry breaking (SSB): the Poincar\'e-invariant vacuum of any relativistic QFT
with an unbroken ${\rm U}(1)$ symmetry;
\item
Zero charge density, but SSB: the Higgs phase of the Standard Model;
\item
Nonzero charge density, but no SSB: a Fermi liquid;
\item
Nonzero charge density and SSB: a superfluid. 
\end{enumerate}

Nevertheless, it is believed that case 3 is realized only for the free Fermi gas: all interacting Fermi liquids end up forming Cooper pairs in the deep infrared and eventually transition to a superfluid or possibly inhomogeneous phase \cite{Shankar:1993pf}.\footnote{Another possibility, in the presence of gapless bosons, is the onset of non-Fermi liquid behavior. We leave aside this currently poorly understood scenario from our discussion.} For instance, experimentally, helium-3 behaves as a degenerate Fermi liquid at temperatures between $\sim$K and $\sim$mK, but at even lower temperatures it turns into a superfluid \cite{vollhardt2013superfluid}. As for systems with bosons only, with some caveats\footnote{There are in fact explicit examples of interacting bosonic theories in $d=3$ that at finite $\mu$ display an emergent fermionic behavior, with bosons satisfying an effective exclusion principle. This is the case of bosonic Chern--Simons theories at large $N$~\cite{Minwalla:2020ysu} (see~\cite{Geracie:2015drf} for earlier work on fermionic Chern--Simons theories displaying bosonic behavior). Even though the coupling of the scalar field to Chern--Simons gauge fields proves crucial in renormalizing the particle spin, we are not aware of a general proof that similar (or other exotic) phenomena cannot occur in theories of interacting scalar fields. See for instance Ref.~\cite{Ciccone:2022zkg} for an example of exotic phase at finite $\mu$ and $T=0$ in $d=2$ QFT (see also~\cite{Thies:2006ti} for a review of earlier works on the same model). In lattice systems, an additional possible phase is provided by the (bosonic) Mott insulator (we thank Sean Hartnoll for remarking this to us).}, 
there is no known example of a state with nonzero charge density that does {\em not} break the corresponding symmetry. 
So, it appears that, at zero temperature, under very general conditions, a nonzero charge density implies spontaneous symmetry breaking. 

This expectation is so ingrained in the way we think about finite density systems, that it is a more or less implicit assumption in much of the recent ``large-charge" CFT literature, starting with the seminal paper \cite{Hellerman:2015nra}.
To appreciate why it is a nontrivial assumption, apart from considering the free Fermi gas case, where it is manifestly violated, one can consider a self-interacting massive complex scalar $\Phi$ with a ${\rm U}(1)$ symmetry.

There, a homogeneous state $| \Psi \rangle$ has nonzero density $J^0 = i \, \Phi^* \overset{\leftrightarrow}{\partial}_t \Phi$ for the ${\rm U}(1)$ charge if and only if
\begin{equation} \label{nonzero density}
\langle \Psi | \Phi^* \overset{\leftrightarrow}{\partial}_t \Phi | \Psi \rangle \neq 0 \; .
\end{equation}
On the other hand, $| \Psi \rangle$ breaks the ${\rm U}(1)$ symmetry if and only if there exists a charged local operator, for instance $\Phi(x)$ itself, with a nonzero expectation value on $| \Psi \rangle$:
\begin{equation} \label{SSB}
\langle \Psi | \Phi | \Psi \rangle \neq 0 \; .
\end{equation}
These two conditions look quite independent, and neither seems to be implying the other. Certainly, there are systems obeying \eqref{SSB} that do {\em not} obey \eqref{nonzero density} (see case 2 above.) Why is it then that all systems obeying  \eqref{nonzero density} happen to obey \eqref{SSB} as well?

At the classical field theory level, there is no mystery: since the density is bilinear in $\Phi$ and $\Phi^*$, to have nonzero density one needs a nonzero $\Phi$, which breaks the symmetry. At the free QFT level also there is  no mystery: at nonzero charge density there is the phenomenon of Bose--Einstein condensation, which implies that the symmetry is spontaneously broken (although it takes some work to prove this last implication \cite{Strocchi:2008gsa}). So, the real question is at the level of the interacting, quantum theory. 

It is important to notice that, with the exception of the somewhat degenerate case of a free theory, there is a control parameter---the chemical potential $\mu$---that can be used to modulate the density. 
Classically, one immediately finds that, if at vanishing $\mu$ the symmetric phase with $\Phi=0$ is stable  and the field there has mass $m$, then for $\mu <m $ the system will stay in that phase, with no charge density and no symmetry breaking, while for $\mu > m $ it will simultaneously develop a nontrivial $\Phi$ and a nontrivial density  $J_0$. 
However, at the quantum level the two operators are two distinct operators, with different quantum numbers, and it is thus a sensible question to ask whether the scales at which they develop a non-zero expectation value, say $\mu=\Lambda_\Phi$ and $\mu=\Lambda_J$, are the same or not.

In this notation,  at tree level one has two, in principle independent, equalities:
\begin{align}
\Lambda_\Phi & = \Lambda_J \equiv \mu_{\rm crit} \\
\mu_{\rm crit} & = m \; .
\end{align}
We may thus ask:  Does the first equality survive at the quantum level? If it does, how is the second corrected? One of our main results will be that, for scalar field theories with generic non-derivative self-interactions, {\em both} equalities survive at one-loop order, with $m$ now being replaced by the physical pole mass of the scalar quanta in the unbroken phase, $\mu_{\rm crit} = m_{\rm pole}$. Moreover, we will find that the same result holds non-perturbatively in the ${\rm O}(N)$ vector model with quartic interactions in $d=3, 4$, at leading order in the large $N$ expansion. We shall also quantify the amount of symmetry breaking as a function of the charge density, by deriving universal scaling relations at low and high density and a strict lower bound. 
As a byproduct, we shall derive the one-loop phonon effective actions for the associated superfluid phases, which are directly related to their equations of state. These results will reproduce and generalize the independent computation of~\cite{Joyce:2022ydd}.

In the following we adopt a path integral approach and pay particular attention to the $i\varepsilon$ terms needed to project onto the ground state of the interacting system at finite $\mu$.  We shall see that the $i\varepsilon$ term projects on a time-independent field configuration (both at zero and at finite density) only in a specific basis of field variables, in which the generalized Lagrangian $\mathcal{L}_{\mu}$ is explicitly $\mu$-dependent and has quadratic terms which are of first order in time derivatives. We shall perform our computations in this basis, so that the properties of the ground state of the system can be extracted from the finite $\mu$ quantum effective potential $V_{\rm eff}$. Moreover, the structure of the poles of the propagators including the finite $\mu$ $i\varepsilon$ term is always such that observables can be computed in Euclidean space, since the Wick rotation is an analytic continuation that does not cross any singularity. When computing the one-loop effective potential we shall therefore compute the one-loop integrals in Euclidean space.
In a separate work~\cite{NPS} we shall consider systems of fermions at finite chemical potential and show how, crucially, an accurate treatment of the $i\varepsilon$ term allows to compute finite $\mu$ quantities such as the free-energy of a Fermi gas using path integral methods. The $\mu$ (in)dependence of the fermionic path integral for small $\mu$ in QCD has been analyzed in~\cite{Cohen:2003kd}; see also~\cite{Dondi:2022zna} for a recent study of the large charge sector of fermionic CFTs in $d=3$ and their infrared phases. Related computations for bosonic systems have been performed previously in Refs.~\cite{Kapusta:1981aa,Bernstein:1990kf,Benson:1991nj,Sharma:2022jio,Brauner:2006xm}. Kapusta~\cite{Kapusta:1981aa} performed a finite-temperature field theory analysis employing the so-called quasi-classical quasi-particle technique that does not fully capture the one-loop corrections in a systematic way, as already noted in the same paper. This computation was later improved by Bernstein, Dodelson and Benson~\cite{Bernstein:1990kf,Benson:1991nj}. The authors considered scalar fields in $d=4$ and formulated the effective potential computation in a way similar to ours, then specializing to the $\lambda \phi^4$ model. The finite temperature computation was also recently reconsidered in Ref.~\cite{Sharma:2022jio}. The effective potential, however, is expressed only as an implicit integral over loop momenta. Our results are in agreement with those of~\cite{Bernstein:1990kf,Benson:1991nj} whenever they overlap. Brauner~\cite{Brauner:2006xm} formulates the calculation of the effective potential for the theory of a complex scalar doublet with $\phi^4$ interactions and internal symmetry $\rm SU(2)\times U(1)$ in $d=4$, and resorts to numerical calculations for the study of its minimum and other properties of interest. Related work has also appeared in the context of pion condensation~\cite{Son:2000xc}, see for instance~\cite{Adhikari:2019mdk,Adhikari:2019mlf,Adhikari:2020ufo}. In particular, Adhikari, Andersen and Kneschke have shown that in chiral perturbation theory the pion condensation transition occurs for a critical chemical potential equal to the pion pole mass at next-to-leading order. On a more formal side, some properties of the relativistic $\lambda \phi^4$ model at finite $\mu$ have been analyzed in the context of axiomatic QFT in Ref.~\cite{Brunetti:2019cax}.

To the best of our knowledge, the integral representations for the finite $\mu$ effective potential and the closed form expressions for its minimum and the related superfluid effective Lagrangian that we obtain, as well as the general statements on symmetry breaking at finite density and the scaling relations for the symmetry breaking scale, are new results that have not appeared before. 

Our discussion on spontaneous symmetry breaking will be limited to theories in which the number of spacetime dimensions is strictly larger than two. The reason for this is the well-known fact that spontaneous symmetry breaking in quantum mechanics and in two dimensional field theories is a subtle concept and requires more care. For instance, the Coleman--Mermin--Wagner theorem~\cite{Coleman:1973ci,Mermin:1966fe} implies that in two-dimensional theories with a Lorentz invariant ground state there is no spontaneous symmetry breaking, at least in the ordinary sense of local order parameters for internal global symmetries. The computation of the effective potential is formally valid also in low dimensions, and some physical quantities can be meaningfully extracted from it, as we shall see in the example of a spinning rigid rotor (Appendix~\ref{app:rotor}). However, in the low-dimensional case the analysis of the effective potential is not enough to make definite statements on spontaneous symmetry breaking, and in our approach this shows up as a breakdown of perturbation theory. (We leave a detailed discussion of these aspects to a future work~\cite{NP}.) Nevertheless, a formal analysis of the quantum mechanical case $d=1$ allows to extract useful physical information and is carried out as a warm-up problem.

Since our paper is rather long and technical, we provide here a roadmap of its structure: 
\begin{itemize}
\item We review in section~\ref{sec:scalar_mu} the derivation of the finite $\mu$ Lagrangian for a complex scalar field $\Phi$ with arbitrary ${\rm U}(1)$-symmetric potential, discussing in detail the role of the $i \varepsilon$ term and the structure of poles in the propagators. In section~\ref{sec:functional_formalism}, we apply functional methods to our system, introducing the formal ingredients that we will use in the rest of the work. In particular, we briefly review the quantum effective action, highlighting the differences due to the presence of a chemical potential $\mu$ and the subtlelties associated with the use of the generating functional $W[j]$ and the effective potential. We then formulate our general question in this language.
 
\item In section~\ref{sec:qm}, as a warm up example, we compute the one-loop effective potential of a quantum mechanical point particle in a central potential. In the related appendix~\ref{app:rotor} we rederive the ground state energy quantization of the spinning rigid rotor in three dimensional space, see eqs.~\eqref{rr1} and \eqref{rr2}.

\item In section~\ref{sec:qftd}, we derive the one-loop effective potential of a complex scalar field with arbitrary ${\rm U}(1)$-invariant, non-derivative self-interactions in generic $d>2$ spacetime dimension. We show explicitly that the expectation that finite density is always accompanied by the spontaneous breaking of the ${\rm U}(1)$ internal symmetry remains valid at the quantum level at one loop (section~\ref{sec:ssb}). We also prove that the critical value of $\mu$ above which the system can support a finite density state is given by the pole mass of the scalar field in the $\mu=0$ theory. Moreover, we derive a universal analytic expression for the one-loop effective action that describes the superfluid phase of these theories and determine the order of the finite density phase transition (section~\ref{sec:super_eft}). We then quantify the amount of symmetry breaking (section~\ref{sec:scaling_relations}) by explicitly deriving universal scaling relations between the charge density and symmetry breaking scale, in the low and high-density limits. These results, together with those of the next section, are the main results of our work.
 
\item In section~\ref{sec:ON}, we study an ${\rm O}(N)$-symmetric theory of $N$ real scalars with quartic interactions in $d=3$ and show that the same conclusions hold non-perturbatively at the leading order of the large-$N$ limit. This theory provides also an explicit example of a model whose finite density dynamics at large density is described by a superfluid effective field theory (EFT) that is {\em not} that of a conformal superfluid. This property is related to the super-renormalizability of the model.
\item In section~\ref{sec:app}, we consider some special cases and compare with previous results in the literature. In particular, we compute (to leading order in $1/n$) the one-loop scaling dimension of the lightest charge $n$ operator  in the ${\rm U}(1)$ theory of a massless complex scalar field with $\phi^6$ interactions in $d=3$, and show that it is in agreement with the numerical result of Ref.~\cite{Badel:2019khk}. 
In addition, we specialize our results of section~\ref{sec:qftd} to the case of a $\phi^4$ potential in $d=4$ and further comment on the connection with Ref.~\cite{Joyce:2022ydd}. In the low density limit we also reproduce the Lee--Huang--Yang relation for the superfluid energy density and compute relativistic corrections to it.
More technical aspects and details are collected in the Appendix.
\end{itemize}

\vspace{.5cm}
\noindent
{\bf Notation and conventions.\\} 
We work in flat spacetime and adopt the mostly minus signature $\eta_{\mu\nu}={\rm diag}(+1, -1, \dots, -1)$ for the metric. We denote by $d$ the number of space-time dimensions.
Throughout the paper $\Phi = (\varphi_{1} +i \, \varphi_{2})/ \sqrt{2}$ will denote a complex scalar field, with real components $\varphi_i$.  We shall use the shorthand notation $\phi= |\Phi| =\sqrt{(\varphi_{1}^{2}+\varphi_{2}^{2})/2}$ for its norm. 
To simplify the notation, we shall assume without loss of generality that $\mu>0$.
By a slight abuse of notation, we will use the symbol $\Q$ to denote both the expectation value of the conserved charge and its volume density, suppressing factors of volume. The correct meaning of the symbol should be clear from the context and volume factors can be reintroduced as desired by dimensional analysis. In this article we assume that the scalar has positive $m^2$.


\section{Complex scalar field at finite chemical potential}
\label{sec:scalar_mu}

We start by reviewing the formulation of a zero-temperature scalar QFT at finite chemical potential, paying special attention to the $i \varepsilon$ terms. We stress that this is crucial to correctly derive the effective potential $V_{\rm eff}$ and compute the observables on the ground state at finite $\mu$.  For the sake of the presentation, we shall mostly focus here on the free theory of a complex scalar field. We shall mention at the end of the section how the discussion generalizes in the presence of an interaction potential.

\subsection{Lagrangian and Hamiltonian formulations}
Let $\Phi(x)$ be a free complex scalar of mass $m$. In Minkowski space with  mostly-minus signature for the metric, its Lagrangian density is
\begin{equation}
\mathcal{L} = (\partial_{\nu}\Phi)^{\dagger} (\partial^{\nu}\Phi) - m^{2} \Phi^{\dagger} \Phi,
\end{equation}
and the $\rm U(1)$ symmetry $\Phi \rightarrow e^{-i \alpha} \Phi$ is manifest. 

The theory can be rewritten in terms of two real scalars $\varphi_i$, $
\Phi = (\varphi_{1} +i \, \varphi_{2})/ \sqrt{2}
$,
with Lagrangian density
\begin{equation}
\mathcal{L} = \dfrac{1}{2} \, \partial_{\nu} \varphi_{i}  \, \partial^{\nu} \varphi_{i} - \dfrac{m^{2}}{2} \varphi_{i}\varphi_{i} \, .
\end{equation}
Under an infinitesimal $\rm U(1)$ transformation, $\delta \Phi = - i \alpha \Phi$, the real scalars transform by an $\rm SO(2)$ rotation,
\be
\delta \varphi_i = \alpha \, \epsilon_{ij} \, \varphi_j  \; ,
\end{equation}
so that the Noether current associated with the symmetry is
\begin{equation}
J^{\nu} = \dfrac{\partial \mathcal{L}}{\partial(\partial_{\nu}\varphi_{i})} \dfrac{\delta \varphi_{i}}{\delta \alpha} = \epsilon_{ij}  \, \partial^{\nu} \varphi_{i}\, \varphi_{j} \; .
\end{equation}

We are interested in studying this system at zero temperature but in the presence of a finite chemical potential for the $\rm U(1)$ charge. To define the system at finite $\mu$ we switch to the canonical formalism and work with a generalized Hamiltonian that includes a chemical potential term, $H_\mu \equiv H - \mu Q$.
The conjugate momenta associated with the real scalar fields are
\begin{equation}
\pi_{i} = \dfrac{\partial\mathcal{L}}{\partial\dot{\varphi_{i}}} = \dot{\varphi_{i}}.
\end{equation}
The canonical Hamiltonian density is readily obtained:
\begin{equation}
\mathcal{H} =\pi_{i} \dot{\varphi_{i}} - \mathcal{L}= \dfrac{1}{2} \pi_i \pi_i  + \dfrac{1}{2} \vec \nabla \varphi_{i} \cdot \vec \nabla \varphi_{i} + \dfrac{m^{2}}{2} \varphi_i \varphi_i  \; ,
\end{equation}
and the generalized Hamiltonian density at finite $\mu$ is 
\begin{equation}
\mathcal{H}_{\mu}= \mathcal{H}- \mu J^{0} = \dfrac{1}{2} \pi_i \pi_i  + \dfrac{1}{2} \vec \nabla \varphi_{i} \cdot \vec \nabla \varphi_{i} + \dfrac{m^{2}}{2} \varphi_i \varphi_i - \mu \, \epsilon_{ij} \, \pi_{i}\varphi_{j} \; .
\label{Hmu}
\end{equation}
The corresponding Lagrangian---the Lagrangian at finite chemical potential---is just the Legendre transform of this, and reads
\begin{equation}
\mathcal{L}_{\mu} = 
\dfrac{1}{2} \, \partial_{\nu} \varphi_{i}  \, \partial^{\nu} \varphi_{i}  - \dfrac{1}{2}(m^{2}-\mu^{2}) \varphi_{i} \varphi_i + \mu \, \epsilon_{ij} \,\dot{\varphi_{i}}\varphi_{j} \; .
\end{equation}
After  integration by parts, this can be equivalently expressed in the matrix form
\begin{equation}
\mathcal{L}_{\mu} = \dfrac{1}{2} 
\begin{pmatrix}
\varphi_{1} &\varphi_{2}
\end{pmatrix}
\cdot K \cdot
\begin{pmatrix}
\varphi_{1}\\
\varphi_{2}
\end{pmatrix},
\end{equation}
with
\begin{equation}
 K=\begin{pmatrix}
- \Box -m^{2}+\mu^{2} & -2\mu \partial_{t} \\
2\mu \partial_{t} & - \Box -m^{2}+\mu^{2} 
\end{pmatrix} \; .
\end{equation}
By looking at the zeroes of the determinant of $K$ in momentum space, one can read off the poles of our fields' propagators. Thinking for the moment only about the small $\mu$ case, $\mu < m$, 
we have positive energy solutions
\begin{equation}
\omega_\pm^{(+)} = \omega_k \mp \mu \;  ,
\end{equation}
where $\omega_k$ is the standard relativistic expression
\be
\omega_{k} =  \sqrt{k^{2}+m^{2}} \; , \qquad k^2\equiv \big|\vec k \big|^2 \; ,
\ee
and negative energy ones, 
\be
\omega_\pm^{(-)} = -\omega_k \mp \mu \; .
\ee

To explain the notation and gain some intuition, consider again our modified Hamiltonian, $\mathcal{H}_{\mu}= \mathcal{H}-\mu Q$. As we will explain in detail, for $\mu < m $ the ground state is still the Poincar\'e invariant vacuum, and the excitations are still the standard Fock states. So, we have particles with charge $q=1$ and anti-particles with charge $q=-1$. Their energies as measured by $\mathcal{H}_{\mu}$ are thus shifted by $\mp \mu$ compared to their standard ones. So, going back to the frequencies above: $\omega^{(+)}_+$ is the positive energy (superscript $(+)$) solution for the positively charged (subscript $+$) particle, and so on.

\subsection{Path integral formulation and $i \varepsilon$ terms}\label{sec:iepsilon}
Consider now the path integral formulation of the theory, in particular that for  time-ordered correlation functions on the ground state of the modified Hamiltonian $\hat H_\mu = \int d^{d-1} x \, \mathcal{\hat{H}}_{\mu}$. We can project onto that state by introducing an appropriate $i \varepsilon$ term in the Hamiltonian path integral. Following the standard procedure, we define the partition function
\begin{equation}
Z(\mu) = \int D\varphi D\pi \, e^{i \int{\rm d}^dx \, I(\pi_{i},\varphi_{i}) }\; ,
\label{piZmu}
\end{equation}
with 
\begin{align}
I(\pi_{i},\varphi_{i}) &= \pi_{i} \dot{\varphi}_{i} -\mathcal{H}_{\mu}+i\varepsilon \mathcal{H}_{\mu} \nonumber \\
&= \pi_{i} \dot{\varphi}_{i} - \dfrac{1}{2}  (1-i \varepsilon) \Big[ \pi_i \pi_i  +  \vec \nabla \varphi_{i} \cdot \vec \nabla \varphi_{i} + m^2 \varphi_i \varphi_i - 2 \mu \, \epsilon_{ij} \, \pi_{i}\varphi_{j} \Big] \; .
\end{align}
Since the exponent in the path integral \eqref{piZmu} is a quadratic polynomial in the momenta, we can play the usual game of solving for the momenta and derive a Lagrangian version of the path integral. After some straightforward manipulations and keeping up to first order in $\varepsilon$, we arrive at
\be
Z(\mu) = \int D\varphi \, \exp \Big\{i \int{\rm d}^dx \, \big(\mathcal{L}_{\mu}(\varphi_{i},\partial \varphi_{i}) + i \varepsilon \, \mathcal{E}_{\mu} (\varphi_{i},\partial \varphi_{i})\big)  \Big\} \; ,
\ee
where  ${\mathcal{L}}_{\mu}$ is the same Lagrangian at finite $\mu$ as above, and
\begin{equation}\label{eq:iepsilon}
 \mathcal{E}_{\mu}(\varphi_{i},\partial \varphi_{i}) \equiv \dfrac{1}{2}\left[\dot{\varphi_{i}}\dot{\varphi}_{i} + 
\vec \nabla \varphi_{i} \cdot \vec \nabla \varphi_{i}    + (m^{2}-\mu^{2}) \varphi_{i} \varphi_{i} \right] \; .
\end{equation}

The Lagrangian $i \varepsilon$ term \eqref{eq:iepsilon} is thus weighted by the Hamiltonian of a free complex scalar with squared mass $(m^{2}-\mu^{2})$. In particular, it does not contain any mixing terms between the two components $\varphi_i$. Reading off its positivity properties is thus straightforward:
\begin{itemize}
\item For $\mu< m$, $\mathcal{E}_{\mu}$ is a positive definite quadratic form in the space of functions over which we are integrating. The path-integral is thus convergent, and, upon a Wick rotation, is  equivalent to the Euclidean one. In particular, in this case our $i \varepsilon$ term is equivalent to the usual one, $i \varepsilon \int \frac12 \varphi_i \varphi_i$.

\item
For $\mu>m$, the mass term in $\mathcal{E}_{\mu}$ is negative definite, which signals that the path integral is not convergent. This, as we will see, is related to a ghost-like instability. In a free  theory there is no cure. With self-interactions instead, this signals that we are expanding about the wrong saddle point.

\item
For $\mu = m$, the mass term in $\mathcal{E}_{\mu}$ vanishes. This makes the zero mode of $\varphi$ a flat direction, which, in a free theory, is associated with the phenomenon of Bose--Einstein condensation.

\end{itemize}

So, in a free theory only $\mu \le m$ is allowed. When we add an interaction potential term $-V_{\rm int}(\varphi)$ to the original Lagrangian, all manipulations above go through unaltered, and the end result is that now $\mathcal{L}_{\mu}$ is supplemented by the same term $-V_{\rm int}(\varphi)$, while the $i \varepsilon$ term includes a $+V_{\rm int}(\varphi)$ piece.
Its positivity properties in the vicinity of $\varphi = 0 $ are thus the same as above. However, when $\varphi = 0$ becomes unstable, that is for $\mu \ge m$, the interaction potential can make $\mathcal{E}_{\mu}$ positive definite about a different, but constant, field configuration $\varphi \neq 0$. This will be the new saddle point one has to expand about, which will lead to SSB.

Repeating the analysis of the previous section, we find that the $i \varepsilon$ term shifts the poles from the real line to the second and fourth quadrants of the complexified frequency plane, for all values of $\mu$. As a result, the analytic continuation from Minkowski to Euclidean space can be performed without crossing any singularity.

\subsection{The effective potential and the ground state}
\label{sec:functional_formalism}

In order to study the properties of the ground state in an interacting QFT it is often convenient to
use functional methods~\cite{Jona-Lasinio:1964zvf,Coleman:1973jx,Jackiw:1974cv}. This approach allows us to extend the semiclassical approximation beyond tree level in a systematic and well-defined way and to include the dynamical effects of external sources. We briefly review this approach in order to highlight the differences introduced by the chemical potential $\mu$ and the role of the $i \varepsilon$ term.

We start from the Lagrangian path-integral representation of the partition function in the presence of sources $j_i(x)$,
\begin{equation}
Z[j_{i};\mu] \equiv \int D\varphi \exp \left\{i \int {\rm d^{d}}x \left(\mathcal{L}_{\mu} [\varphi]+ j_{i}(x)\varphi_{i}(x) + i \varepsilon\, \mathcal{E}_{\mu} [\varphi] \, \right)\right\} \; . \label{Z(j,mu)}
\end{equation}
We treat $\mu$ as a constant parameter on the same footing as the other couplings.\footnote{We shall see, however, that, differently from the mass and the self-interaction couplings, the chemical potential $\mu$ does not need to be renormalized.} $Z[j_{i} ; \mu] $ is a functional of the sources and its functional derivatives generate all the time-ordered Green's functions of $\hat{\varphi}_{i}$ in the presence of the sources $j_{i}(x)$. It is often more convenient to work with the generating functional of connected Green's functions,
\begin{equation}
W[j;\mu] = - i \log Z[j;\mu].
\end{equation}
The so-called \emph{classical field} is defined as the expectation value of $\hat{\varphi}(x)$ in the presence of the source $j(x)$:
\begin{equation}
\label{eq:phiclassical}
\varphi_{cl}(x) = \dfrac{\delta W[j;\mu]}{\delta j(x)}= \langle \Omega \vert \hat{\varphi}(x) \vert \Omega \rangle_{j, \mu} \; .
\end{equation}
The quantum effective action $\Gamma[\varphi_{cl};\mu]$ is defined through the Legendre transform
\begin{equation}
\Gamma[\varphi_{cl};\mu] = W[j;\mu] - \int {\rm d}^{d}x \, j(x) \varphi_{cl}(x),
\end{equation}
where $j(x)$ is understood as a functional of $\varphi_{cl}(x)$, through the inverse of equation~\eqref{eq:phiclassical}. $\Gamma[\varphi_{cl};\mu]$ generates one-particle-irreducible (1PI) Green's functions, as it can be proved by taking appropriate functional derivatives. From the definition of the Legendre transform it follows that
\begin{equation}
\dfrac{\delta \Gamma[\varphi_{cl};\mu]}{\delta \varphi_{cl}(x)}= - j(x).
\end{equation}
In particular, the expectation value of $\hat{\varphi}(x)$ for vanishing external source must be a stationary point of the effective action, \textit{i.e.}~it obeys $\delta \Gamma[\varphi_{cl};\mu]/\delta \varphi_{cl}(x) =0$.

The quantum effective action admits a loop expansion, and, perhaps more importantly, a derivative expansion.  The lowest order in the latter corresponds to constant field values, and, in that limit, the quantum effective action reduces to just an effective potential term, $- \int {\rm d}^d x \,V_{\rm eff}(\varphi_{cl};\mu)$, which generates correlation functions with vanishing external momenta. 
At tree level, the effective potential is just the ordinary potential $V(\varphi_{cl};\mu)$, while at all orders it has a representation in terms of a path-integral over quantum fluctuations about ${\varphi}_{cl}$:
\be \label{1PIPI}
e^{-i \int V_{\rm eff}(\varphi_{cl}; \mu)} = \int_{\rm 1PI} D \delta \varphi \, e^{i \int \big( \mathcal{L}_{\mu}[\varphi_{cl} + \delta \varphi] + i \varepsilon\, \mathcal{E}_{\mu}[\varphi_{cl} + \delta \varphi]  \big)} \; ,
\ee
where, in a diagrammatic expansion, the path-integral is restricted to 1PI diagrams only.
In the absence of sources, {\em if} the ground state is translationally invariant, it must have an expectation value for $\hat \varphi$ that minimizes the effective potential,
\be
\langle \Omega | \hat \varphi(x)  | \Omega \rangle_\mu = \bar {\varphi} \; , \qquad \frac{\partial V_{\rm eff}(\varphi_{cl}; \mu)}{\partial \varphi^i_{cl}} 
\bigg|_{\bar \varphi} = 0  .
\ee

All this is absolutely standard, but
now we come to two technical subtleties that, though important for our study, once understood can be safely ignored:
\begin{enumerate}
\item 
The path-integral expression \eqref {1PIPI} makes no sense for certain values of $\varphi_{cl}$: if the quadratic terms in the expansion of $\mathcal{E}_{\mu}$ about a certain $\varphi_{cl}$ are {\em not} positive definite, then the (perturbative) path integral does not converge. This is physically associated with the fact that, even in the presence of a suitable source $j(x)$, the state with that $\varphi_{cl}$ as expectation value for $\hat \varphi$ is unstable. This is easier to understand in pictures than in words (or formulae)---see fig.~\ref{bottle}. Technically, even in cases when $W[j ; \mu]$ is well defined for all sources $j(x)$,  its Legendre transform might not exist for all classical fields $\varphi_{\rm cl}(x)$.
\begin{figure}[t!]
\begin{center}
\includegraphics[width=5in]{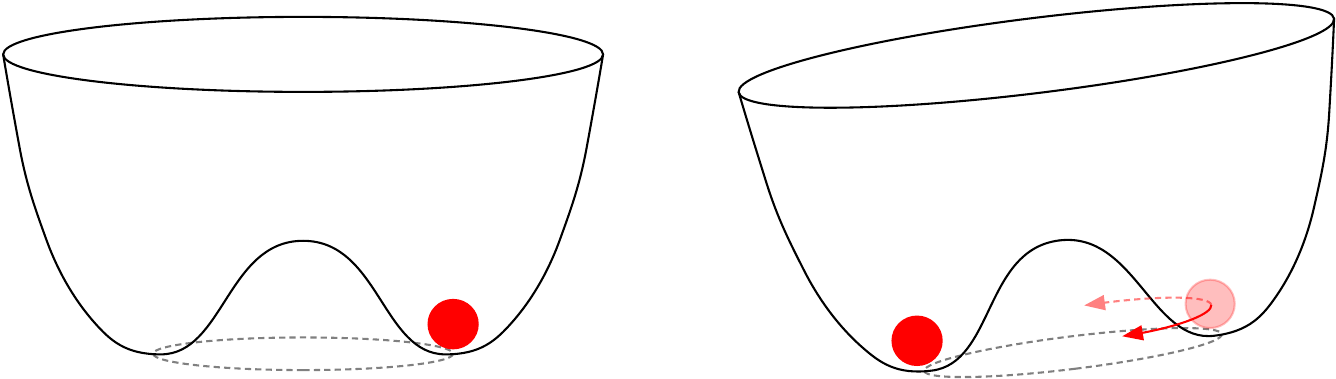}
\end{center}
\caption{\label{bottle} \it \small The instability discussed in the text. If one starts with a $\rm U(1)$-invariant potential with SSB (left), adding a linear source (right) can destabilize the equilibrium position if this is ``on the wrong side." As a result, the  effective potential is not formally defined in the region inside the valley of minima, because there are no values for the source that can lead to such expectation values.}
\end{figure}

The responsible thing to do would then be to compute $W[j ; \mu]$ rather than the effective action or the effective potential. 
However, in the usual perturbative computations of the effective potential one uses the standard $i \varepsilon$ prescription, \textit{i.e.}, $+i \varepsilon $ at the denominator of Feynman propagators. Then, the subtlety just alluded to shows up as an imaginary part for the effective potential in the ``forbidden" range of $\varphi_{cl}$, which then one interprets, correctly, as an instability.

We will take this pragmatic approach, keeping in mind that, to be safe, one 
should approach the minima of the effective potential from the stable side---typically, larger field values, in absolute value.

\item
To study whether or not the system features SSB, we will study the expectation value of the field doublet $\varphi_i$. Now, in our path-integral formulation of the partition function, eq.~\eqref{Z(j,mu)}, the only ingredient that breaks the $\rm SO(2)$ symmetry is the coupling to the source $j_i(x)$.
So, apparently, at zero source, whatever we compute from the partition function must be symmetric, and so there cannot be SSB.

The zero source limit, however, is more delicate than that. Recall that, for any value of the source, the partition function yields the expectation value of the field via eq.~\eqref{eq:phiclassical}. The fact that the source is the only symmetry breaking ingredient in the partition function implies that, at least for sources that are constant in spacetime, the expectation value thus obtained will be aligned with the source, with a symmetry-preserving coefficient
\be \label{finite vev}
\langle \vec \varphi \, \rangle_{\mu, j} = v\big(  | \vec j \,| ; \mu \big) \, \frac{\vec j}{| \vec j \,|} \; , \qquad \vec j(x) = {\rm const.}
\ee
So, from this viewpoint it is clear that SSB is equivalent to the statement that the function $v$ has a nonzero limit for $\vec j$ going to zero: an external source or perturbation, however small, will determine the {\it direction} in which the symmetry is broken, but the amount of breaking remains finite even for arbitrarily small sources, and it is thus an intrinsic property of the system. This approach to SSB is, in fact, quite physical, and resolves the apparent paradox alluded to above.\footnote{It is sometimes called the Bogolyubov approach \cite{Strocchi:2008gsa}.}

At the technical level, this means that SSB is associated with a non-analyticity of the generating functional $W[j; \mu]$ at zero $\vec j$, because an analytic behavior consistent with the symmetries, $W = {\rm const} + |\vec j|^2 + {\cal O}(|\vec j|^4)$, would imply vanishing derivatives at zero $\vec j$, and thus vanishing expectation values for the field $\vec \varphi$. In particular,  a finite limit for eq.~\eqref{finite vev} requires
\be \label{W of j}
W[j; \mu] \sim | \vec j \,| \; , \qquad \vec j \to 0 \; .
\ee

Once again, we can bypass this subtlety and just study the effective potential---in particular, its minima: approaching a minimum of the effective potential is equivalent to sending a source to zero. If that minimum corresponds to a nonzero field value, the system features SSB.

\end{enumerate}
We are now in a position to phrase our physical QFT question in the functional framework just described.

\subsection{The fundamental question}
Consider our $\rm U(1)$-invariant complex scalar QFT.
Its effective potential $V_{\rm eff}(\varphi; \mu)$, at generic values of the chemical potential $\mu$, must be $\rm U(1)$-invariant. So, as far as $\varphi$ is concerned, it can only depend on the absolute value $\phi = |\vec \varphi|$,
\be
V_{\rm eff} = V_{\rm eff}(\phi = |\vec \varphi|; \mu) \; .
\ee

From $V_{\rm eff}$ we can compute many physical properties of the ground state at finite $\mu$. In particular, the expectation value of $\vec \varphi$ must have absolute value
\be
\big| \langle \vec \varphi \, \rangle_\mu  \big| = \phi_{\rm min}(\mu) \; ,
\ee
where $\phi_{\rm min}(\mu)$ is a, possibly $\mu$-dependent, minimum of $V_{\rm eff}$,
\be
\frac{\partial V_{\rm eff}(\phi; \mu)}{\partial \phi} \bigg|_{\phi_{\rm min}(\mu)} = 0.
\ee

On the other hand, 
from the Hamiltonian path-integral definition of the partition function at vanishing source, see eqs.~\eqref{piZmu} and~\eqref{Hmu}, and from its relationship to the effective potential, it follows that the ground state's charge density is
\begin{equation}\label{eq:chargevev}
Q(\mu) \equiv \langle   \hat{J}^{0}   \rangle_{\mu} =-i \dfrac{{\rm d} }{{\rm d} \mu}\log Z(\mu)= -\dfrac{{\rm d} V_{\rm eff}(\phi_{\rm min}(\mu);\mu)}{{\rm d}\mu} = -\dfrac{\partial V_{\rm eff}(\phi ;\mu)}{\partial \mu} \bigg|_{\phi_{\rm min}(\mu)} \; ,
\end{equation}
where we used that working at zero source is equivalent to working at $\phi_{\rm min}$ and that $V_{\rm eff}$ is stationary there. 
(To avoid clutter, we are suppressing spacetime volume factors. If needed, these can be reinstated by dimensional analysis.)

{\em If} at zero chemical potential there is no SSB  and the vacuum is the usual Poincar\'e invariant one for relativistic field theories, then:
\be
\phi_{\rm min}(0) = 0 \; , \qquad Q(0) = 0 \; .
\ee
Imagine now turning on a (positive) chemical potential. What happens? We like to think of the chemical potential as a control parameter for the charge density. But, in fact, in free theory, or at the classical level for an interacting theory, {\em nothing} happens for a finite range of $\mu$, specifically, for $\mu < m$, where $m$ is the mass of our scalar particles: the ground state is still the Poincar\'e  invariant vacuum, with zero charge density and unbroken $\rm U(1)$ symmetry.

In free theory, when $\mu$ reaches $m$ the system develops  a charge density {\em and} SSB at the same time, and for $\mu > m$ it is unstable. So, in bosonic free theory, the chemical potential is not a good control parameter at all. Things are better behaved in the classical interacting theory: for $\mu \ge m$ the system exhibits both a charge density {and} SSB, and the charge density $Q(\mu)$ and symmetry breaking scale $\phi_{\rm min}(\mu)$ are both increasing functions of $\mu$.

So, our fundamental question is whether and how this story gets modified at the quantum level in the interacting theory. In particular:

{\em 1) Does $Q(\mu)$ remain zero for a finite range of $\mu$'s and, if so, what is the critical $\mu$ above which it becomes nonzero, and

2) Is there a range of $\mu$'s for which $Q(\mu)$ is nonzero but the symmetry is unbroken ($\phi_{\rm min}(\mu) = 0$)?}

\section{Quantum Mechanics: the point particle in a central potential}
\label{sec:qm}

As a warm up, we can explore all these ideas in quantum mechanics. Let us consider a point particle moving in a plane in a central potential. In the presence of a large enough 
chemical potential for the angular momentum, the naive semiclassical configuration in which the particle sits at rest at the center of the potential becomes unstable. At this point, the system develops both a nonzero angular momentum  and a symmetry-breaking expectation value for the position. In fact, as is well known, there is no SSB in quantum mechanics at the non-perturbative level. In our formalism, this property will show up as a breakdown of perturbation theory for such symmetry-breaking expectation value.

To see all this, consider first the case of a quadratic potential, corresponding to a two-dimensional harmonic oscillator. The generalized Hamiltonian is:\footnote{We use the letter $m$ to denote the frequency of the oscillator in order to have a uniform notation with the field theoretical case. This parameter should not be confused with the mass of the point particle, 
which is taken to be unity.}
\begin{equation}
H_\mu  = \dfrac{1}{2} \, \vec p \, ^2 + \dfrac{m^{2}}{2} \vec q \, ^2 + \mu \, \epsilon_{ij} q_i p_j \; .
\end{equation}
This Hamiltonian can be diagonalized via a time-independent canonical transformation,
\begin{equation}
\begin{split}
&Q_{1}=\sqrt{\dfrac{m}{2}} q_{1} + \sqrt{\dfrac{1}{2m}} p_{2} \; , \quad Q_{2}=\sqrt{\dfrac{m}{2}} q_{2} + \sqrt{\dfrac{1}{2m}} p_{1} \; , \\
&P_{1}=-\sqrt{\dfrac{m}{2}} q_{2} + \sqrt{\dfrac{1}{2m}} p_{1} \; , \quad P_{2}=-\sqrt{\dfrac{m}{2}} q_{1} + \sqrt{\dfrac{1}{2m}} p_{2} \; , 
\end{split}
\end{equation}
leading to
\begin{equation}
H_\mu = \dfrac{1}{2}(m+\mu)\big(P_{1}^{2}+Q_{1}^{2} \big) + \dfrac{1}{2}(m- \mu)\big(P_{2}^{2}+Q_{2}^{2} \big) \; .
\end{equation}
In these variables it becomes transparent that:
\begin{itemize}
\item
For $\mu < m $ we have two independent harmonic oscillators, with frequencies $\omega_\pm = m \pm \mu$. The ground state corresponds to vanishing occupation numbers for both of them, regardless of the value of $\mu$ (within this range). Since the canonical transformation that we performed does not depend on $\mu$ either, then this ground state is just the standard $\mu = 0$ ground state of the system. That is, as anticipated, nothing happens for $\mu < m$.
\item 
For $\mu = m$ the system becomes degenerate: the Hamiltonian of the second ``oscillator" becomes identically zero. This degeneracy reflects the fact that states of arbitrary non-zero angular momentum are all mapped to $\mu = m$ in the absence of interactions. It will be lifted by interaction terms.
\item 
For $\mu > m$ we have a {\em ghost} instability: the Hamiltonian of the second oscillator is negative definite. 
\end{itemize}
To study the fate of this instability, we now go back to the original $(\vec q, \vec p \,)$ variables and consider anharmonicities in the potential. 

For definiteness, consider a quartic potential:
\begin{equation} \label{QM quartic}
V(q)=  \dfrac{m^{2}}{2} \vec q \, ^2 +  \dfrac{\lambda}{4}\big(\vec q \, ^2\big)^{2}.
\end{equation}
The Hamiltonian can no longer be diagonalized with a simple canonical transformation, and in order to study the properties of the ground state it is useful to switch to the Lagrangian formalism and study the effective potential.
The finite $\mu$ Lagrangian is 
\begin{equation}
L_{\mu}= \dfrac{1}{2} \dot{\vec q} \, ^2- \dfrac{(m^{2}-\mu^{2})}{2} \vec q \, ^2 - \dfrac{\lambda}{4}\big(\vec q \, ^2\big)^{2} + \mu \, \epsilon_{ij} \dot q_i q_j \; ,
\end{equation}
and we now apply \eqref{1PIPI} at one-loop: we expand the action to quadratic order about a generic but constant classical position $\vec q_{cl} =(\bar q_1, \bar q_2)\,$,
\begin{equation}
\begin{split} \label{QMS2}
&S_{(2)}[ \delta \vec q \, , \vec q_{\rm cl}]=-\dfrac{1}{2} \int {\rm d}t \;
\delta \vec q  \,  (t)\cdot
K(\vec q_{cl} ) \cdot
\delta \vec q \, (t)
\; , \\
&K(\vec q_{cl} )=\begin{pmatrix}
\partial_{t}^{2} +m^{2}-\mu^{2} + 3 \lambda \bar{q}_{1}^{2} +\lambda \bar{q}_{2}^{2} &-2\mu \partial_{t} + 2 \lambda \bar{q}_{1} \bar{q}_{2} \\[0.5em] 
 2\mu \partial_{t} + 2 \lambda \bar{q}_{1} \bar{q}_{2} & \partial_{t}^{2} + m^{2} - \mu^{2} + \lambda \bar{q}_{1}^{2} + 3 \lambda \bar{q}_{2}^{2}
\end{pmatrix} \; ,
\end{split}
\end{equation}
and evaluate the gaussian path integral that yields the one-loop correction to the effective potential, $\Delta V^{\rm 1L}_{\rm eff}$. 

Following standard functional methods \cite{Weinberg:1996kr},
\begin{align}
e^{- i \int \! dt \, \Delta V^{\rm 1L} _{\rm eff}[\vec q_{cl}; \mu]} & = {\int}D\delta q(t) \, e^{i \,S_{(2)}[\delta \vec q \, ]} \\
& = \big({\rm Det} \, K \big)^{-1/2} \\
& = \exp \Big\{-\frac12 \int \! dt \int \! \frac{d\omega}{2 \pi}  \log \det \tilde K(\vec q_{cl} ) \Big\} \; ,
\end{align}
where `${\rm Det}$' is a functional determinant, `$\det$' an ordinary one, and $\tilde K$ the Fourier-space version ($\partial_t \to -i \omega$) of the $K$ matrix in \eqref{QMS2}.

So, the one-loop correction to the effective potential reads 
\begin{equation}
\Delta V^{\rm 1L}_{\rm eff}(\vec {q}_{cl}; \mu)=- \dfrac{i}{2} \int \dfrac{\rm d \omega}{2\pi} \log\left[\left( \omega^{2}-\omega_{-}^{2} \right) \left( \omega^{2}-\omega_{+}^{2} \right) \right] \; ,
\end{equation}
where $\omega_{\pm}$ are the poles of the $\delta \vec q$ propagators, which depend on the classical radial distance $r \equiv |\vec q_{cl} |$ and on $\mu$:
\begin{equation}
\omega_{\pm}^{2} (r; \mu) = m^{2}+\mu^{2} +2 \lambda r^{2}\pm \sqrt{4 m^{2}\mu^{2}+8 \lambda \mu^{2} r^{2}+ \lambda^{2} r^{4}} \; .
\end{equation}
Computing the integral in Euclidean space,  renormalizing away an $r$-independent and $\mu$-independent zero-point energy, and including the tree-level contributions, we arrive at the final one-loop result
\begin{equation} \label{QM loop}
V_{\rm eff}(r ; \mu)= \dfrac{1}{2}(m^{2}-\mu^{2}) r^{2} + \dfrac{\lambda}{4} r^{4}+ \dfrac{1}{2} \big( \omega_{+} (r; \mu)+\omega_{-}(r; \mu) \big) .
\end{equation}

Let us start by asking for what values of $\mu$ the origin is a stable (or metastable) configuration. This is determined by the sign of the second $r$-derivative of $V_{\rm eff}$ at $r=0$. We have
\be
\frac{\partial^2 V_{\rm eff}(r ; \mu)}{\partial r^2} \bigg|_{r=0}
 = m^{2}-\mu^{2}+\dfrac{2 \lambda}{m} .
 \ee
 So, the origin is stable for
 \be
 \mu^2 < m^2 + \frac{2 \lambda}{m} \; ,
 \ee
and unstable otherwise.

Since in this system there is no wave-function renormalization at one loop, the second derivative of the potential at the origin and at $\mu = 0$ happens to be the `pole mass' $m^2_{\rm pole}$---the renormalized energy of the first excited states at vanishing chemical potential. So, the origin is stable for 
\be
 \mu^2 < m_{\rm pole}^2  \; ,
 \ee
 and unstable otherwise.

Within this stability range, we can ask whether the system does feature a nonzero angular momentum at finite $\mu$. This is determined by the $\mu$-derivative of $V_{\rm eff}$ at the origin. However, we have
\be
V_{\rm eff}(0; \mu) = m  \; ,
\ee
and so its $\mu$-derivative vanishes for all $\mu$'s.

We thus reach the same conclusion as in the classical (or free quantum) theory: nothing happens until $\mu$ crosses a critical value, given by the pole mass. Beyond that value, the system develops both a symmetry breaking average position and a nonzero angular momentum. 
However, this is where things go wrong in perturbation theory, as anticipated above. Let us start with the classical (\textit{i.e.}, tree-level) limit. For $\mu > m$ the expectation value of the position and of the angular momentum are determined by the minimum of the classical potential \eqref{QM quartic}, and read
\be
r^2_{\rm min} = \dfrac{\mu^{2}-m^{2}}{\lambda} \; , \qquad J = -\frac{\partial V}{\partial \mu} \bigg|_{r_{\rm min}} = \dfrac{\mu (\mu^{2} - m^2)}{\lambda }
\qquad \qquad \mbox{(tree level)} \; .
\ee
Now, it so happens that the 1-loop correction in \eqref{QM loop} is singular at $r = r_{\rm min}$, making it impossible to correct the value of $r_{\rm min}$ order by order in  perturbation theory. To see this, it is enough to notice that for $\mu > m$, the pole frequency $\omega_-$ displays a singularity at $r = r_{\rm min}$:
\be
\omega_- \sim \sqrt{r- r_{\rm min}} \; , \quad r \to r_{\rm min} \qquad \qquad (\mu > m) \; .
\ee
As explained around eq.~\eqref{W of j}, some form of non-analyticity is to be
expected for SSB to take place. However, the one above for $\omega_-$ is too strong. One can check that it corresponds to a one-loop generating functional scaling as
\be
W[j; \mu] \sim \sqrt{| \vec j \,|} \; , \qquad \vec j \to 0 \; ,
\ee
which is not regular at zero source. We take this as a sign of a breakdown of perturbation theory.

As mentioned above, the fact that perturbation theory breaks down in this case is consistent with the fact that there should not be SSB in quantum mechanics. We devote another paper to study in some detail this fact and the analogous one in $d=2$ with functional methods \cite{NP}. 
As well known, these obstructions to SSB do not apply for QFT in $d > 2$, which we study next. However, as a check of our methods, in Appendix~\ref{app:rotor} we also generalize the QM analysis to the case of a quantum mechanical rigid rotor, rederiving the well-known quantization condition $E_J= {J(J+1)}/{2I}$  for its energy.

\section{QFT in $d > 2$ dimensions}
\label{sec:qftd}

We now consider the case of a complex scalar field in $d > 2$ dimensions, where $d$ includes both space and time and will be taken as an arbitrary parameter. 

We consider the case in which the tree-level potential includes the mass term and arbitrary ${\rm U}(1)$-invariant self-interactions, so that at finite $\mu$ the kinetic term and the tree-level potential are, respectively,
\begin{equation}\label{eq:phiself}
\begin{split}
& \mathcal L_{\mu,{\rm kin}} = (\partial_{\nu}\Phi)^{\dagger}  (\partial^{\nu}\Phi) + i\mu \,\Phi^{\dagger}\overset{\leftrightarrow}{\partial}_0 \Phi \\
&V(\phi;\mu)= (m^2-\mu^2) \phi^2 + V_{\rm int}(\phi) \; .
\end{split}
\end{equation}
The only assumption we impose for the interactions is regularity, that is, that $V_{\rm int}(\phi)$ admits a Taylor expansion around $\phi=0$ that starts at quartic or higher order.

\begin{figure}[t]
\begin{center}
\includegraphics[width=10cm]{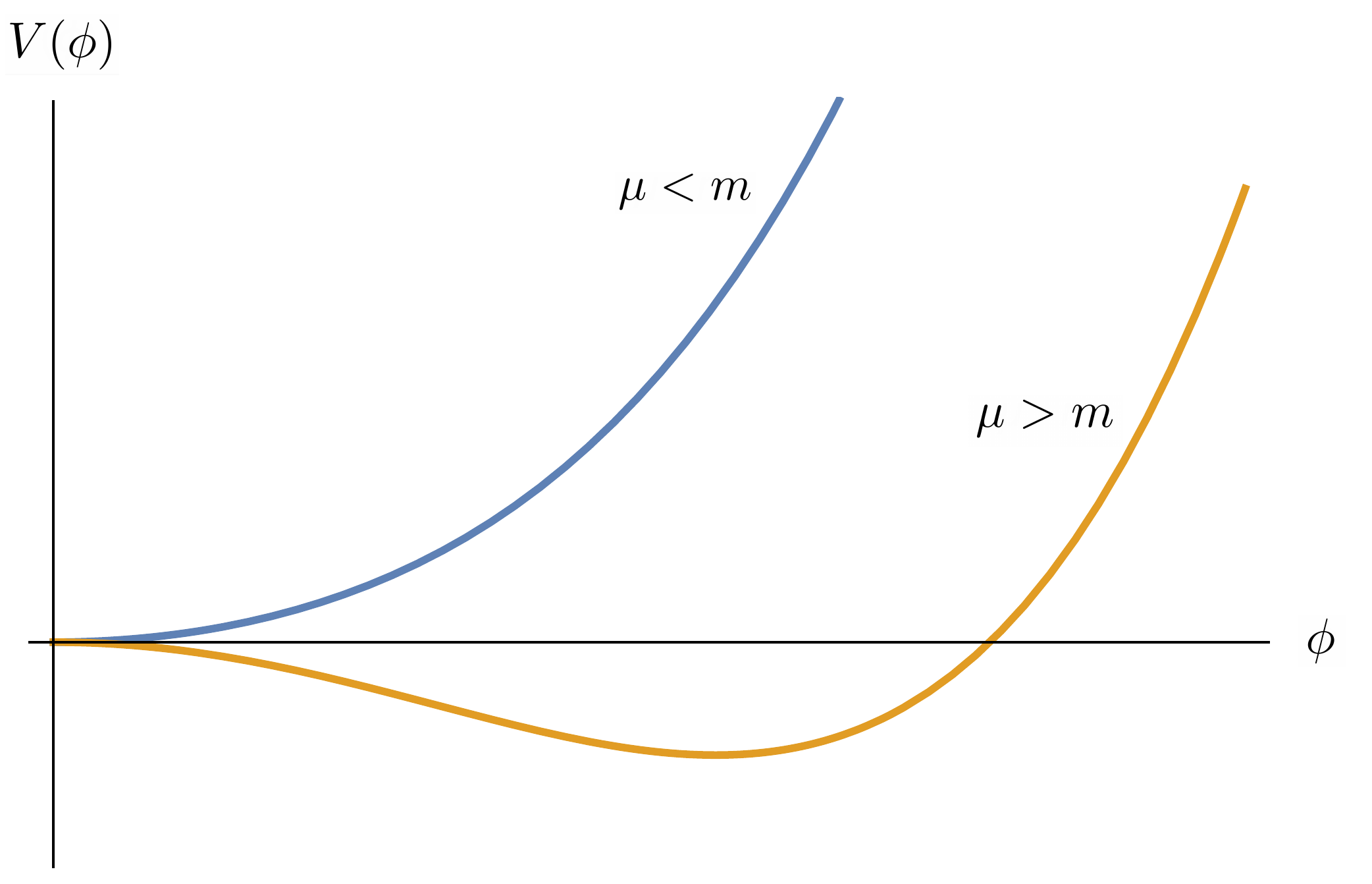}
\caption{\label{full potential} \small \it For simplicity we restrict to potentials $V(\phi;\mu)$ that, as a function of $\phi \equiv |\Phi|$, only feature one minimum. For $\mu < m$ the minimum is at the origin, while for $\mu > m $ it is at a nonzero value of $\phi$.}
\end{center}
\end{figure}

For simplicity we restrict to interaction potentials that are growing functions of $\phi$, so  that the full potential $V(\phi;\mu)$ is minimized at the origin for $\mu < m$ and develops a single symmetry-breaking minimum for $\mu > m$, as depicted in fig.~\ref{full potential}. More general interactions can be considered, in which case the full potential can have more extrema. 

\subsection{Free scalar for $\mu < m$}\label{sec:freescalarmu}
Let us first treat the familiar case of free massive bosons, obtained by setting $V_{\rm int}(\phi)=0$. The system is unstable for $\mu>m$, and we want to check that for $\mu < m$ the system is equivalent to a system of free bosons at zero density, which implies that the only non-trivial case is then the degenerate case $\mu = m$.

Carrying out the procedure detailed in Appendix~\ref{app:det}, the path integral in Euclidean space yields
\begin{equation}
\log Z(\mu) = - \dfrac{i}{2}{\rm Vol} \int \dfrac{{\rm d}^{4}p}{(2\pi)^{4}}\left[ \log  \left(p^{2} +m^{2}+\xi_{0}^{2}+ 2 \xi_{0} p_{0}\right)  + \log\left(p^{2} +m^{2}+\xi_{0}^{2}- 2 \xi_{0} p_{0}\right) \right] \; ,
\end{equation}
where $\xi_0=i\mu$ and we neglected a $\mu$-independent additive constant.
Let us focus on the first term and use dimensional regularization to compute:
\begin{equation}\label{eq:iplus}
I_{+}=\int \dfrac{{\rm d}^{d}p}{(2\pi)^{d}} \log \left(p^{2} +m^{2}+\xi_{0}^{2}+ 2 \xi_{0} p_{0}\right).
\end{equation}
Completing the square for the time component of the momentum,
\begin{equation}
I_{+}= \int \dfrac{{\rm d}^{d}p}{(2\pi)^{d}} \log \left((p_0+\xi_{0})^{2}+ |\vec p \,|^2 +m^{2}\right),
\end{equation}
and shifting the integration variable $p_{0}$ we end up with 
\begin{equation}
I_{+}= \int \dfrac{{\rm d}^{d}\tilde{p}}{(2\pi)^{d}} \log \left(\tilde{p}^{2}+m^{2} \right)
\; ,
\end{equation}
which, clearly, does not depend on $\mu$. 

Analogous manipulations apply to the second term in $Z(\mu)$ above, and we thus reach the conclusion
\be
Z(\mu < m) = Z(0) \; ,
\ee 
which, upon deriving w.r.t.~$\mu$, implies that the charge density vanishes for all $\mu < m$. Although not manifest in our derivation, this condition is crucial to make the computation in Euclidean space well defined---as emphasized in section~\ref{sec:iepsilon}, the $i \varepsilon$ terms are such that for $\mu > m $ the Euclidean path integral does not converge.

Another technical subtlety is that in the derivation above we had to rely on a momentum-shift in the imaginary direction. 
In Appendix~\ref{app:muindependence} we provide an alternative derivation of this result which does not need such a shift.

\subsection{Effective potential at finite $\mu$}

We now introduce self-interactions and consider the ${\rm U}(1)$-invariant Lagrangian~\eqref{eq:phiself}.

The analysis of the $i \varepsilon$ term~\eqref{eq:iepsilon} suggested that, at tree level, for $\mu >m$ the semiclassical configuration $\vec \varphi(x)\equiv 0$ is unstable. As we discussed in section~\ref{sec:functional_formalism}, with our choice of field variables and generalized Hamiltonian, in order to find the correct semiclassical ground-state configuration at finite $\mu$, it is sufficient to consider the effective potential and study its minima.

The effective potential can be computed by functional methods~\cite{Coleman:1973jx,Jackiw:1974cv}. By working in an arbitrary number of dimensions and using the general results derived in Appendix~\ref{app:det},
the one-loop effective potential at finite $\mu$ for our case is
\begin{equation}
\label{eq:veff_alpha}
V_{\rm eff}^{(1)}(\varphi;\mu) = \dfrac{1}{2} \int \dfrac{{\rm{d}}^{d}p}{(2\pi)^{d}} \log \left[(p^{2}+M^{2})^{2} - 4 (p \cdot \xi)^{2} - g^2 \right],
\end{equation}
where $p$ is the Euclidean momentum and we defined
\begin{equation}\label{eq:M2g2_general}
\begin{split}
& \xi \equiv ( i \mu, \vec{0} \, ), \\
&M^{2} = M^{2}(\phi; \mu) \equiv\dfrac{1}{4\,\phi}\Big(V'(\phi; \mu) + \phi \,V''(\phi;\mu)\Big) , \\
&g^2= g^2(\phi) \equiv \dfrac{1}{16\,\phi^2}\Big(V'(\phi; \mu) - \phi \,V''(\phi; \mu)\Big)^2,
\end{split}
\end{equation}
where primes denote derivatives with respect to $\phi$. 
In what follows, when needed, we shall denote by $g$ the positive square root of $g^2$. Notice that, as proved in Appendix~\ref{app:det} in full generality, the combination $(M^2+\mu^2)$ is $\mu$ independent, while $g^2$ is $\mu$-independent and vanishes for $\phi=0$. 

Sometimes we shall specialize to the physical dimensions $d=3$ and $d=4$, or consider the explicit case of a complex scalar with $ \phi^\alpha$ interactions,
\begin{equation}
V_{\rm int}(\phi)= \lambda \phi^\alpha,
\end{equation}
in which case
\begin{equation}
\begin{split}
& M^{2}= m^{2}-\mu^{2}+\dfrac{\alpha^2}{4} \lambda \, \phi^{\alpha-2} = m^{2}-\mu^{2}+\dfrac{\alpha}{(\alpha-2)}g,\\
& g^2 =  \lambda^2 \, \dfrac{\alpha^2 (\alpha -2)^2}{16} \, \phi^{2\alpha-4}.
\end{split}
\end{equation}
This includes as particular cases the renormalizable $\lambda \phi^4$ model in $d=4$ and $\lambda \phi^6$ model in $d=3$, as we shall discuss in what follows,  reproducing and generalizing some results previously obtained in other works. 

One might worry that the argument of the $\log$ in eq.~\eqref{eq:veff_alpha} can become negative for some values of $\phi$ and $p^2$, generating an imaginary part for the effective potential. This is indeed the case, in general, at the left of its minimum. It is straightforward to check that for $p=0$ the argument of the $\log$ is positive for $\phi>\phi_0$, where $\phi_0$ is the minimum of the tree level, finite $\mu$ potential, but can be negative for $\phi$ close to $\phi_0$, but smaller than it. This follows by noticing that
 \begin{equation}
M^4-g^2 = \dfrac{V'V''}{4\phi} 
\; ,
\end{equation}
which changes sign precisely at the minimum of $V$.
Stronger positivity properties hold for nonzero Euclidean momenta, ensuring that the one-loop effective potential is real for $\phi>\phi_0$, as expected from the general considerations of section~\ref{sec:iepsilon}.

There are two useful ways of organizing the calculation, which consist in expanding the logarithm either in powers of $g$ (subsection~\ref{sec:gexpansion}) or in powers of $\mu$ (subsection~\ref{sec:muexpansion}).\footnote{A third way to organize the computation allows to rewrite the effective potential in closed form. This is achieved by first integrating on frequency, similarly to the QM case~\eqref{QM loop}, and then computing the spatial momentum integral in terms of $\vec{p}^{\,2}$, up to a shift. For integer $d$, the resulting integrals are Abelian integrals: of the (hyper-)elliptic type for even $d$; expressible in terms of elementary functions for odd $d$. We find the approaches described in the main text better suited for our purposes.}

\subsection{Expanding in powers of $g$}
\label{sec:gexpansion}

We use the decomposition
\begin{equation}
(p^{2}+M^{2})^{2} - 4 (p \cdot \xi)^{2} = A\cdot B,
\end{equation}
with
\begin{equation}
\begin{split}
& A= p^{2}+M^{2}+2 p \cdot \xi = (p+\xi)^{2}+ M^{2} - \xi^{2}, \\
& B= p^{2}+M^{2}-2 p \cdot \xi = (p-\xi)^{2}+ M^{2} - \xi^{2},
\end{split}
\end{equation}
and write the $\log$ as
\begin{equation}
\begin{split}
\log \left[(p^{2}+M^{2})^{2} - 4 (p \cdot \xi)^{2} - g^2  \right] &= \log A + \log B + \log \left(1-\dfrac{g^2}{A\cdot B} \right) = \\
&=\log A + \log B - \sum_{n=1}^{\infty}\dfrac{1}{n} \left(\dfrac{g^2}{A\cdot B} \right)^{n}.
\end{split}
\end{equation}
In order to compute the integral in~\eqref{eq:veff_alpha} we introduce a Feynman parameter through the identity 
\begin{equation}
\dfrac{1}{A^{n}B^{n}} = \int_{0}^{1} {\rm d}x \dfrac{[x(1-x)]^{n-1}}{(x A + (1-x) B)^{2n}} \dfrac{\Gamma(2n)}{\Gamma(n)^{2}} \; ,
\end{equation}
and complete the square in the denominator as: 
\begin{equation}
x A + (1-x) B = (p + (2x-1) \xi )^{2} + M^{2} - (2x-1)^{2} \xi^{2}.
\end{equation}
Evaluating the integrals in an arbitrary number of dimensions we obtain the general result for the dimensionally regularized one-loop effective potential in $d$ dimensions:
\begin{multline}
\label{eq:Veff_g_d}
V_{\rm eff}^{(1)}(\phi;\mu) = -\dfrac{\Gamma(-d/2)}{(4\pi)^{d/2}}(M^{2} + \mu^{2})^{d/2}
\\
 - \dfrac{1}{2} \sum_{n=1}^{\infty} \dfrac{\Gamma(2n-d/2)}{(4\pi)^{d/2}\Gamma(n)^{2}} \dfrac{g^{2n}}{n} \int_{0}^{1} {\rm d}x \dfrac{[x(1-x)]^{n-1}}{(M^{2} + (2x-1)^{2}\mu^{2})^{2n-d/2}} \; .
\end{multline}

In odd dimensions, $d=2k+1$, at one loop there are no logarithmic divergencies and the one-loop effective potential is regular for $d\rightarrow 2k+1$. Therefore, in this case the dimensionally regularized expression~\eqref{eq:Veff_g_d} is finite and no counterterms are needed in the minimal subtraction scheme. The series can be resummed in closed form for any odd dimension in terms of a hypergeometric function. 
For example, in $d=3$ we arrive at the result
\begin{equation}
\label{eq:Veff_g_3}
V_{\rm eff}^{(1)}(\phi;\mu)\Big\vert_{d=3} = -\dfrac{1}{6\pi}(M^{2} + \mu^{2})^{3/2} - \dfrac{g^2}{16\pi} \int_{0}^{1}{\rm d}x \,\sqrt{y} \;{}_2F_1\left[\begin{array}{c}
\frac{1}{4},~~\frac{3}{4}\\
2
\end{array}\Bigg\vert 4g^2 \, x(1-x) \, y^2 \right]
\; ,
\end{equation}
where 
\begin{equation}
y=\dfrac{1}{M^2+(1-2x)^2 \mu^2},
\end{equation}
and $M^2,g^2$ are given by eq.~\eqref{eq:M2g2_general} for a generic potential.

In even dimensions $d=2k$, more care is needed as some of the terms in~\eqref{eq:Veff_g_d} are divergent for $d\rightarrow 2k$ and need to be renormalized. In $d=4$, all the terms with $n\geq 2$ in the series are regular. The divergent contribution is 
\begin{equation}\label{eq:poles}
V_{\rm eff}^{(1)}(\phi;\mu)\Big\vert_{d=4,{\rm div}}=\dfrac{g^2+(M^2+\mu^2)^2}{16\pi^2} \dfrac{1}{(d-4)},
\end{equation}
which is $\mu$ independent. Moreover, for local UV theories with polynomial potentials the divergencies can always be canceled by the addition of local countertems, as expected on general grounds. The one-loop divergencies can be $\mu$ dependent in $d=6$ or larger. This is related to the appearance of derivative local counterterms in the $\mu=0$ theory, as we discuss in Appendix~\ref{app:divergencies}. Nothing is lost, however: as we shall argue on general ground and check in explicit examples, the countertems needed at finite $\mu$ match exactly those required to renormalize the UV theory at $\mu=0$.

In $d=4$, by renormalizing the divergencies in the $\overline{\text{MS}}$ scheme and resumming the $n\geq 2$ terms of the series, we obtain
\begin{equation}
\label{eq:Veff_g_4}
\begin{split}
V_{\rm eff}^{(1)}(\phi;\mu)\Big\vert_{d=4} = &- \dfrac{1}{64\pi^2} \left[ 4g^2 + 3 (M^2+\mu^2)^2 \right] + \dfrac{1}{32\pi^2} \left[g^2 + (M^2+\mu^2)^2 \right]\log\left( M^2+\mu^2 \right) \\
&+ \dfrac{g^2}{16\pi^2} \dfrac{M}{\mu}\arctan\left(\dfrac{\mu}{M}\right) 
- \dfrac{g^{2}}{256\pi^{2}} \int_{0}^{1}{\rm d}x \,y \;{}_3F_2\left[\begin{array}{c}
1,~~1,~~\frac{3}{2}\\
2,~~3
\end{array}\Bigg\vert y \right]
\; ,
\end{split}
\end{equation}
where
\begin{equation}
y= \dfrac{4 g^2 \, x (1-x)}{(M^{2}+(1-2x)^{2}\mu^{2})^{2}},
\end{equation} 
and $M^2,g^2$ are given by eq.~\eqref{eq:M2g2_general} for a generic potential. The $\overline{\text{MS}}$ renormalization scale $\bar{\mu}$ is set to $1$ for simplicity, but can be reintroduced on dimensional grounds.

\subsection{Finite density, symmetry breaking, and the critical value of $\mu$}
\label{sec:ssb}

The expressions we just derived for the effective potential at finite $\mu$ are particularly useful in analyzing the relationship between finite density and spontaneous symmetry breaking. 

In odd dimensions $d=2k+1$, at one loop there are no logarithmic divergencies and the one-loop effective potential is regular for $d\rightarrow 2k+1$. From the previous expression and the facts that $g^2=0$ and $(M^{2}+\mu^{2})=m^2$ for $\phi=0$ (see Appendix~\ref{app:det} for a general derivation), it follows that
\begin{equation}
V_{\rm eff}^{(1)}(\phi=0;\mu)\Big\vert_{d\; {\rm odd}} = -\dfrac{\Gamma(-d/2)}{(4\pi)^{d/2}}m^{d},
\end{equation}
for arbitrary interactions. In particular, $V_{\rm eff}^{(1)}(\phi=0;\mu)$ is $\mu$ independent. It follows that in the unbroken phase in which $\varphi=0$ the system cannot have finite charge density --- see eq.~\eqref{eq:chargevev}. From this result we can conclude that, in general, in a system describing a complex scalar at finite density in odd dimension, the global ${\rm U}(1)$ symmetry is always spontaneously broken. 
That is: finite density is always accompanied by spontaneous symmetry breaking, not only at the classical level but also at one loop.

The case of even dimension $d=2k$ requires more care due to the presence of logarithmic divergencies. For $d=4$, renormalizing the theory in the $\overline{\text{MS}}$ scheme for an arbitrary potential --- eq.~\eqref{eq:Veff_g_4} --- and using again the facts that $g^2=0$ and $(M^{2}+\mu^{2})=m^2$ for $\phi=0$, we obtain
\begin{equation}
V_{\rm eff}^{(1)}(\phi=0;\mu)\Big\vert_{d=4}= -\dfrac{3}{64\pi^2} m^4 + \dfrac{1}{32\pi^2} m^4 \log(m^2),
\end{equation}
which is manifestly $\mu$ independent. Again, we conclude that for complex scalar fields, at one loop and in four dimensions finite density for a ${\rm U}(1)$ charge is always accompanied by spontaneous symmetry breaking. 

We stress that there is no ambiguity in this conclusion related to the renormalization scheme, since the theory can be renormalized at $\mu=0$ and no additional counterterms are needed to renormalize the finite $\mu$ theory. This is true even in the case of non-renormalizable field theories, where an infinite number of counterterms is needed to renormalize the theory at arbitrary loop level.\footnote{As usual, only a finite number of counterterms is necessary at a fixed loop order.} The computation of the Coleman--Weinberg potential at finite $\mu$ allows anyway to make definite low-energy predictions for finite density properties of the system, since the counterterms are determined independently of the infrared deformation induced by $\mu$.

\label{sec:criticalmu}

Having established the relationship between finite density and symmetry breaking, an interesting question to address is: what is the critical value of $\mu$ above which the system can support a finite density state?

The analysis of Ref.~\cite{Joyce:2022ydd} hinted that at one loop the critical value $\mu_{\rm crit}$ coincides with the pole mass of the scalar in the $\mu=0$ theory, assuming $m^2_{\rm pole}>0$. This suggestive result was obtained by studying the consistency of the low-energy effective theory for the superfluid phonons in an explicit $\lambda \phi^4$ model. Demanding stability and subluminality of the phonon perturbations, one finds that the theory is well-behaved only for $\mu^2 > m^2_{\rm pole}$. 
A full understanding, however, can only be obtained by studying the UV theory with the inclusion of the radial mode. This is what we shall do in this section, by analyzing the finite $\mu$ one-loop effective potential, for arbitrary interaction potential.

Let us work at one loop and denote by $V_{\rm eff}$ the finite $\mu$ effective potential including both the tree-level and loop contributions.
From eq.~\eqref{eq:chargevev} we have $\Q= - \frac{\rm d}{{\rm d}\mu} V_{\rm eff}(\phi_{\rm min};\mu)$, so that:
\begin{equation}\label{eq:mucrit}
J>0 \iff - \frac{\rm d}{{\rm d}\mu} V_{\rm eff}(\phi_{\rm min};\mu) >0.
\end{equation}
Given the relationship with SSB, the critical value of $\mu$ corresponds to the limit $\phi_{\rm min}\rightarrow 0^+$. 
We can then expand $V_{\rm eff}(\phi_{\rm min};\mu)$ around $0$:
\begin{equation}
V_{\rm eff}(\phi_{\rm min};\mu) = V_{\rm eff}(0;\mu) + V^{'}_{\rm eff}(0;\mu) \phi_{\rm min} + V^{''}_{\rm eff}(0;\mu) \dfrac{\phi_{\rm min}^2}{2} + \dots,
\end{equation}
where primes denote derivatives with respect to $\phi$.
The constant term is $\mu$-independent as we proved in the previous section, so it drops out of eq.~\eqref{eq:mucrit}. In dimensions $d=3$ and $4$, and for $\rm U(1)$ invariant interaction potential, it is easy to show from eqs.~\eqref{eq:phiself},~\eqref{eq:Veff_g_3} and~\eqref{eq:Veff_g_4} that $V^{'}_{\rm eff}(0;\mu) =  0$. This follows simply from the property that $g^2, \partial_\phi g^2, \partial_\phi M^2$  all vanish at $\phi=0$, and is a consequence of the $\rm U(1)$ invariance of the theory.
Therefore, in the limit $\phi_{\rm min}\rightarrow 0^+$ the condition~\eqref{eq:mucrit} becomes:
\begin{equation}
- \frac{\rm d}{{\rm d}\mu}\left( V^{''}_{\rm eff}(0;\mu) \dfrac{\phi_{\rm min}^2}{2} \right)>0.
\end{equation}
Using the fact that $\frac{\rm d}{{\rm d}\mu}\phi^2_{\rm min}>0$, for $\phi_{\rm min}\rightarrow 0^+$  we arrive at the condition:
\begin{equation}
-  V^{''}_{\rm eff}(0;\mu)>0.
\end{equation}
As a final step we separate the contributions from tree-level and one loop. At tree level one has simply $V^{''}(0;\mu)=m^2-\mu^2$. At one loop, using again from eqs.~\eqref{eq:Veff_g_3} and~\eqref{eq:Veff_g_4}, the property that $g^2, \partial^2_\phi g^2$ are vanishing for $\phi=0$, and the property that $M^2+\mu^2$ is $\mu$-independent, it follows that $V^{(1)}_{\rm eff}{}^{''}(0;\mu)\equiv V^{(1)}_{\rm eff}{}^{''}(0;\mu=0)$. Since there is no wave-function renormalization at one-loop in the theory, this is nothing else than the one-loop contribution to the pole mass of the complex scalar in the $\mu=0$ theory. 
Therefore we arrive at the final result:
\begin{equation}\label{eq:mucrit2}
\mu^2 > V^{''}_{\rm eff}(0;0)= m^2_{\rm pole}.
\end{equation}
This proves in general at one loop that $\mu_{\rm crit}^2= m^2_{\rm pole}$, as suggested from the stability analysis of the low-energy effective theory.

Notice that at the technical level, the condition~\eqref{eq:mucrit} corresponds to $\partial_{X}P(X)>0$ in the superfluid effective theory. However, in the low-energy effective theory the relationship between this condition and the pole mass is obscured. Moreover we stress that the condition on $\mu$ derived in the effective theory is a necessary condition for spontaneous symmetry breaking, but the fact that it is also sufficient is apparent only thanks to the study of the effective potential.

\subsection{Expanding in powers of $\mu$ and the superfluid EFT}
\label{sec:muexpansion}

Alternatively, to evaluate \eqref{eq:veff_alpha} we can use the decomposition:
\begin{equation}
(p^{2}+M^{2})^{2} - g^{2} = C\cdot D,
\end{equation}
with
\begin{equation}
\begin{split}
& C= p^{2}+M^{2}-g , \\
& D= p^{2}+M^{2}+g,
\end{split}
\end{equation}
and write the $\log$ as: 
\begin{equation}
\begin{split}
\log \left[(p^{2}+M^{2})^{2} - g^{2} - 4 (p \cdot \xi)^{2} \right] &= \log C + \log D + \log \left(1-\dfrac{4 (p \cdot \xi)^{2}}{C\cdot D} \right) = \\
&=\log C + \log D - \sum_{n=1}^{\infty}\dfrac{1}{n} \left(\dfrac{4 (p \cdot \xi)^{2}}{C\cdot D} \right)^{n}.
\end{split}
\end{equation}
As before, to compute the integral in~\eqref{eq:veff_alpha} we introduce a Feynman parameter through the identity:
\begin{equation}
\dfrac{1}{C^{n}D^{n}} = \int_{0}^{1} {\rm d}x \dfrac{[x(1-x)]^{n-1}}{(x C + (1-x) D)^{2n}} \dfrac{\Gamma(2n)}{\Gamma(n)^{2}},
\end{equation}
and complete the square in the denominator as:
\begin{equation}
x C + (1-x) D = p^{2} + M^{2} + g - 2g \, x.
\end{equation}
Evaluating the integrals in an arbitrary number of dimensions by using the results summarized in Appendix~\ref{app:integrals}, we obtain an alternative general expression for the dimensionally regularized one-loop effective potential in $d$ dimensions:
\begin{equation}
\begin{split}
\label{eq:Veff_mu_d}
V_{\rm eff}^{(1)}(\phi;\mu) = &- \dfrac{1}{2} \dfrac{\Gamma(-d/2)}{(4\pi)^{d/2}}(M^{2}-g)^{d/2} - \dfrac{1}{2} \dfrac{\Gamma(-d/2)}{(4\pi)^{d/2}}(M^{2}+g)^{d/2} \\
&- \dfrac{1}{2} \sum_{n=1}^{\infty} (-1)^{n}  \left(\dfrac{\Gamma(n+1/2) \Gamma(n-d/2)}{(4\pi)^{d/2}\Gamma(n)^{2}\Gamma(1/2)}\right) \dfrac{(2\mu)^{2n}}{n} \int_{0}^{1} {\rm d}x \dfrac{[x(1-x)]^{n-1}}{(M^{2}+g-2g\, x)^{n-d/2}}.
\end{split}
\end{equation}
Similarly to the case of eq.~\eqref{eq:Veff_mu_d}, in odd dimensions $d= 2k+1$ this result is finite and no counterterms are needed in the minimal subtraction scheme.
The series can again be resummed, but we shall not do so for the moment, as the integral in ${\rm d}x$ cannot be computed in closed form in general. 

In even dimensions $d= 2k$, as before, more care is needed as some of the terms in~\eqref{eq:Veff_mu_d} are divergent for $d \rightarrow 2k$ and renormalization is required. 
We carry out the explicit computation in $d=4$, where all the terms with $n\geq3$ in the series are regular.  As expected by consistency, summing up the divergent contributions we find again
\begin{equation}
V_{\rm eff}^{(1)}(\phi;\mu)\Big\vert_{d=4,{\rm div}}=\dfrac{g^2+(M^2+\mu^2)^2}{16\pi^2} \dfrac{1}{(d-4)},
\end{equation}
so that the counterterms are the same as those previously mentioned.
By renormalizing the divergencies in the $\overline{\text{MS}}$ scheme we find:
\begin{equation}
\label{eq:Veff_mu_4}
\begin{split}
V_{\rm eff}^{(1)}(\phi;\mu)\Big\vert_{d=4} = & \, \dfrac{1}{384 \pi^2 g^3}
\Bigg[-18 g^5 +6 g \mu^4 M^4 -2 g^3 \left(8 \mu ^4+9 M^4+18 \mu^2 M^2\right) \\
&+3 \left(M^2-g\right)^2 \left(2 g^3-2 g^2 \mu ^2+2 g \mu^4+\mu^4 M^2\right) \log \left(M^2-g\right)\\
&+3 \left(M^2+g\right)^2 \left(2 g^3+2 g^2 \mu ^2+2 g \mu^4-\mu^4
M^2 \right) \log \left(M^2 +g \right) \Bigg]\\
&- \dfrac{1}{32\pi^2} \sum_{n=3}^{\infty} (-1)^{n}  \dfrac{\Gamma(n+1/2) \Gamma(n-2)}{\Gamma(n)^{2}\Gamma(1/2)} \dfrac{(2\mu)^{2n}}{n} \int_{0}^{1} {\rm d}x \dfrac{[x(1-x)]^{n-1}}{(M^{2}+g-2g\, x)^{n-2}},
\end{split}
\end{equation}
where the renormalization scale $\bar{\mu}$ is set to $1$ for simplicity.

\label{sec:super_eft}

This result allows us to derive the Goldstone low-energy effective action that describes the superfluid phase of the theory and, correspondingly, the one-loop equation of state for such a superfluid. In this way we shall generalize the independent result of Ref.~\cite{Joyce:2022ydd}, and reproduce a result of Ref.~\cite{Badel:2019khk} derived using a completely different approach.

As argued in~\cite{Son:2002zn} and easily seen from symmetry arguments, the low-energy quantum effective action for superfluid phonons, at leading order in derivatives, takes the form
$ \Gamma_{\rm eff}\left[X\right] = P(X)$, where $X=(D_{\nu} \pi)(D^{\nu} \pi)$. The covariant derivative $D_{\nu} \pi= \partial_{\nu} \pi+\mu \delta_{\nu}^{0}$ acts non-linearly on $\pi$, being associated with a non-linearly realized global symmetry. 
Assuming that the effective action is extremized for a constant $\pi$ configuration, as dictated by the $i\varepsilon$ term, we end up with the relationship 
\begin{equation}
\label{eq:PXfromVeff}
P(X) \xleftrightarrow[\pi = \rm const]{X=\mu^2} - V_{\rm eff}(\phi_{\rm min}(\mu);\mu).
\end{equation}

At one loop, it is sufficient to set $\phi$ to the value at which the tree-level potential is minimized, denoted by $\phi_{0}$, thanks to the fact that $V'(\phi_{0})=0$.
As proved in general in Appendix~\ref{app:det}, one has
\begin{equation}\label{eq:gmin}
M^2\Big\vert_{\rm min} = g\Big\vert_{\rm min} = \dfrac{V''(\phi_{0})}{4} \equiv g_{\rm min} \ .
\end{equation}
Consider then  the ${\rm d}x$ integral in the general expression~\eqref{eq:Veff_mu_d},
\begin{equation}
I(n,d;\phi)\equiv \int_{0}^{1} {\rm d}x \dfrac{[x(1-x)]^{n-1}}{(M^{2}+g-2g\, x)^{n-d/2}}.
\end{equation} 
At $\phi= \phi_{0}$ the integrand simplifies and we can compute $I(n,d;\phi_{0})$ analytically in terms of the Euler beta function (or equivalently in terms of Gamma functions):
\begin{equation}
\begin{split}
I(n,d;\phi_{0})&=\left(\dfrac{1}{2g_{\rm min}}\right)^{n-d/2} \int_{0}^{1} {\rm d}x \, x^{n-1} (1-x)^{d/2-1} \\
&= \left(\dfrac{1}{2g_{\rm min}}\right)^{n-d/2} B\left(n,\dfrac{d}{2}\right) = \left(\dfrac{1}{2g_{\rm min}}\right)^{n-d/2} \dfrac{\Gamma\left(n\right)\Gamma\left(d/2\right)}{\Gamma\left(n+d/2\right)}.
\end{split}
\end{equation} 

Using this identity, the relation~\eqref{eq:PXfromVeff} and resumming the series, we can compute the Goldstone effective action for arbitrary potential. 
In odd dimensions $d=2k+1$, from eq.~\eqref{eq:Veff_mu_d} it follows
\begin{equation}
\begin{split}
P(X)\Big\vert_{d\, \rm odd}= & - V(\phi_{0};\mu=X^{1/2}) \\
&+ \left( -1 \right)^{(d+1)/2}\dfrac{\pi}{2 \, \Gamma\left(d/2+1\right)} \left( \dfrac{g_{\rm min}}{2\pi} \right)^{d/2} \;{}_2F_1\left[\begin{array}{c}
\frac{1}{2},~~-\frac{d}{2}\\
\frac{d}{2}
\end{array}\Bigg\vert -\frac{2X}{g_{\rm min}}\right]
\; ,
\end{split}
\end{equation}
where $g_{\rm min}$ should be seen as a function of $\mu$ upon the substitution $\mu=X^{1/2}$.
The hypergeometric function of interest can be expressed in a more familiar way in terms of polynomials, square roots and hyperbolic functions. In particular, in $d=3$ it takes the explicit form:
\begin{equation}\label{eq:PX_3d}
\begin{split}
P(X)\Big\vert_{d=3}= & - V(\phi_{0};\mu=X^{1/2}) \\
&+ \dfrac{1}{48\pi} \left[ \left(5 g_{\rm min} + 4X\right) \sqrt{2 g_{\rm min}+4X} + 3 \left(\dfrac{g_{\rm min}^2}{\sqrt{X}}\right) \,{\rm arcsinh}\left(\sqrt{\dfrac{2X}{g_{\rm min}}}\right) \right].
\\ 
\end{split}
\end{equation}
The inverse hyperbolic function ${\rm arcsinh}(z)$ can be equivalently expressed as $\log(z + \sqrt{z^2+1})$.

On the other hand, in $d=4$, from the $\overline{\text{MS}}$ renormalized result of eq.~\eqref{eq:Veff_mu_4} it follows that
\begin{equation}\label{eq:PX_4d}
\begin{split}
P(X)\Big\vert_{d=4}= & - V(\phi_{0};\mu=X^{1/2}) \\
&+ \dfrac{1}{192\pi^2} \Bigg( 
- 4 X^2 + \left(X^2 +2 g_{\rm min} X + 2 g_{\rm min}^2 \right) \left(9 - 6\log(2 g_{\rm min})\right)\\
&
\hspace{50pt} -\dfrac{5X^3}{2g_{\rm min}} \;{}_3F_2\left[\begin{array}{c}
1,~~1,~~\frac{7}{2}\\
4,~~5
\end{array}\Bigg\vert -\dfrac{2X}{g_{\rm min}} \right]
\; 
 \Bigg).\\ 
\end{split}
\end{equation}
This result matches exactly the independent computation of~\cite{Joyce:2022ydd} in the $\lambda \phi^4$ theory in $d=4$, where $m^2_{\rm eff}= 2g_{\rm min}= 2(X - m^2)$, and provides an additional non-trivial consistency check.

From these results we see that the one-loop contribution to the $P(X)$ takes a universal form in a fixed number of spacetime dimensions and depends only on the curvature $g_{\rm min}$ of the classical finite $\mu$ potential at its minimum, given by eq.~\eqref{eq:gmin}. Correspondingly, the same observation holds for the one-loop equation of state of a zero-temperature relativistic superfluid, which is obtained by setting $X=\mu^2$ and identifying $P(\mu^2)$ with the pressure $p$.

The results~\eqref{eq:PX_3d} and~\eqref{eq:PX_4d} can be used to derive the universal behavior of the one-loop free energy at the finite density phase transition, determining in particular its order. Denote $\Delta = \mu^2-m_{\rm pole}^2$. We are interested in the behavior of the free energy $f(\mu)=P(\mu^2)$ for $\Delta \rightarrow 0^+$. For smooth and regular potential, the tree-level contribution is analytic around $\Delta=0$, for every $d$. Moreover, the term in $f(\mu)$ of order $\Delta^0$ is a constant independent of $\mu$, equal to (minus) the effective potential in $\phi_{\rm min}=0$ as computed in section~\ref{sec:ssb}, and just corresponds to the cosmological constant contribution (computed for $\mu=0$). The phase transition is therefore of second or higher order. The non-analytic terms originate from the one-loop contribution, which depends only on $\bar{X}= \mu^2 =  m_{\rm pole}^2 + \Delta $ and $g_{\rm min} = c_1 \Delta +  c_2 \Delta^2 + \dots $. Expanding for $\Delta\rightarrow 0^+$ (and setting the renormalization scale $\bar{\mu}=m_{\rm pole})$ we find that 
\begin{equation}\label{nonanalyticities}
\begin{split}
&f(\mu)\Big\vert_{d=3}= -\dfrac{c_1^2}{32\pi \, m_{\rm pole}}\Delta^2  \log\left( \Delta\right) + \mathcal{O}(\Delta^3 \log\left( \Delta\right))+ {\rm analytic\; in\;} \Delta, \\
&f(\mu)\Big\vert_{d=4}= -\dfrac{\sqrt{2} c_1^{5/2}}{15\pi^2 \, m_{\rm pole}}\Delta^{\frac{5}{2}}  + \mathcal{O}(\Delta^{\frac{7}{2}})+ {\rm analytic\; in\;} \Delta. \\
\end{split}
\end{equation}
The leading non-analyticity has therefore a universal behavior in $d=3,4$, with coefficient $c_1=1$ in the presence of a quartic coupling. The phase transition is of second (third) order respectively, since the free energy as a function of $\mu$ has singular second (third) derivative for $\Delta\rightarrow 0^+$.

\section{Quantifying spontaneous symmetry breaking}

\label{sec:scaling_relations}

We have shown that, at least at one loop, a system of scalars at finite density for some internal charge necessarily breaks the corresponding ${\rm U}(1)$  symmetry. However, this statement is not particularly meaningful unless we can quantify, or put a lower bound on the amount of symmetry breaking. After all, there could exist a parametric limit for which the charge density remains constant while all the physical effect of symmetry breaking go to zero.

An obvious candidate to quantify SSB is $\phi_{\rm min}$, the expectation value of $\Phi$ itself, and in particular its relationship to the charge density. At tree level one has
\be
Q = 2 \mu\,  \phi_{\rm min}^2  \qquad \qquad \mbox{(tree level)} \; ,
\ee
and one might wonder whether at loop level such a relationship survives, or perhaps gets corrected in some universal way. However, beyond one loop $\phi$ itself is not particularly meaningful from a physical standpoint---for instance, it is subject to wave-function renormalization. It would be better to find a more direct characterization of the symmetry breaking scale, one that is directly related to observable quantities. 
It is possible to do so by considering the superfluid effective theory, whose general one-loop form was discussed in section~\ref{sec:super_eft}.

Let us consider the case in which the $\rm U(1)$ global symmetry is linearly realized in the UV theory at $\mu=0$, so that $m^2_{\rm pole}>0$.
We know from our previous results that for $\mu^2> m^2_{\rm pole}$ the system is in the superfluid phase and we can consider the low-energy quantum effective action for the superfluid phonons $P(X)$, where $X= \partial_{\nu}\psi \partial^{\nu}\psi$.\footnote{The distinction between the classical and the quantum effective action for a superfluid is irrelevant up to one-loop order in dimensional regularization~\cite{Joyce:2022ydd}. To all orders, the quantum effective action is related to the equation of state of the relativistic superfluid and provides a physical definition of the observable symmetry breaking scale.} Expanding  the effective action in terms of background $\bar{\psi}=\mu t$ and phonon perturbations $\pi(x)$, one finds the quadratic Lagrangian
\be
S_{(2)} = \frac12 \int {\rm d}^d x \,{ f^{d-2}_\pi}\bigg[ \frac{\dot \pi^2}{c_s^2} -  \big(\vec\nabla \pi \big)^2 \bigg] \; ,
\ee
where
\be \label{fpi and cs}
f_\pi^{d-2} (\mu) = 2 P'(\mu^2) \; , \qquad c_s^2(\mu) = \frac{P'(\mu^2)}{2 P''(\mu^2) \mu^2 + P'(\mu^2)}
\ee

The quantity $f_\pi $, which has units of energy, can be taken as a concrete and physical definition of the symmetry breaking scale, in analogy with the pion decay constant in the QCD chiral Lagrangian. In fact, at tree level it coincides with $2 \phi_{\rm min}^2$. Once the normalization of $\pi(x)$ is fixed, for instance by demanding that it be an angular variable of period $2\pi$ (with apologies for the inconsistent usage of `$\pi$'), $f_\pi$ is completely unambiguous, at the non-perturbative level. It can be thought of as a measure of the static rigidity of the ground state, in the sense that it controls the zero-frequency limit of the Goldstone response function.
We shall refer to it as `the symmetry breaking scale'. 

To relate $f_\pi$ to the charge density, we notice that the Noether current associated with the ${\rm U}(1)$ (shift) symmetry of the effective action $P(X)$ is
\begin{equation}\label{eq:phonon_current_1}
J^{\nu}=2 P'(X) \partial^{\nu}\psi ,
\end{equation}
so that on the background $\bar{\psi}=\mu t$ we have 
\be
\Q(\mu)=2 P'(\mu^2) \mu \; .
\ee
From \eqref{fpi and cs} we thus have
\begin{equation}\label{eq:fpi}
f_\pi^{d-2} (\mu)= Q(\mu)/ \mu.
\end{equation}
We now have to eliminate $\mu$ in favor of $Q$. 

Before we try do so for general $Q$, recall that there is a threshold value for $\mu$, the pole mass, below which there is no charge density and no SSB. So, when $\mu$ crosses that threshold but is still very close to it, the charge density is very small and the above relationship simply becomes
\be \label{f(Q)}
f_\pi (Q) \simeq \bigg(\frac{Q}{m_{\rm pole}}\bigg)^\frac{1}{d-2} \qquad \mbox{for } Q \to 0 \; .
\ee
This is a universal, non-perturbative prediction for the symmetry breaking scale in a very dilute superfluid made up of scalar bosons with physical mass $m_{\rm pole}$.

For larger values of $Q$, the most useful way we have found to eliminate $\mu$ from \eqref{eq:fpi} is to derive with respect to $Q$ and use the above relationships involving derivatives of $P$ w.r.t.~$\mu^2$. After straightforward algebra we find the ODE
\be \label{eq:dfpi}
(d-2) \, \dfrac{{\rm d}\log f_\pi}{{\rm d}\log \Q}  = \left(1-c_s^2(Q)\right),
\ee
where $c_s$ is  a function of $Q$ through $\mu$. Such an ODE is to be supplemented by the boundary condition \eqref{f(Q)}. 
The only solution  is thus
\begin{equation} \label{f(Q) general}
f_\pi^{d-2}(\Q) = \left(\dfrac{\Q}{m_{\rm pole}}\right) \exp\left(-\int_0^Q\frac{c_s^2(Q')}{\Q'}{\rm d}Q'\right).
\end{equation}
The integral is always convergent thanks to the low-$\Q$ behavior of $c_s^2$ (see Appendix~\ref{app:details}),
\begin{equation}
c_s^2(\Q)= \dfrac{1}{4m_{\rm pole}^3 P''(m_{\rm pole}^2)} \Q+ \dots
\end{equation}
The small $\Q$ expansions of $f_\pi$ and $\mu$  thus are
\begin{equation}\label{eq:low\Q_scaling}
f_\pi^{d-2}(Q)= \dfrac{\Q}{m_{\rm pole}} - \dfrac{1}{4 m_{\rm pole}^4 P''(m_{\rm pole}^2)} \Q^2 + \dots,
\end{equation}
\begin{equation}\label{eq:low\Q_mu}
\mu (Q)= m_{\rm pole} + \dfrac{1}{4m_{\rm pole}^2 P''(m_{\rm pole}^2)} \Q+ \dots,
\end{equation}
where we used \eqref{eq:fpi}.

Like the leading order term discussed above, the next-to-leading one is also universal, but it involves a new independent parameter, which we can take to be ${\rm d}c_s^2 / {\rm d}\Q$ evaluated at $\Q=0$. This is certainly an observable quantity, but with perhaps a less familiar interpretation. In the $ \phi^4$ model it is determined by the quartic coupling or, equivalently, the scattering length. The derivation of eq.~\eqref{eq:low\Q_scaling} is completely general, valid beyond the one-loop order, if we replace $m_{\rm pole}$ with $\mu_{\rm crit}$, the critical value of the chemical potential. We expect the identification of these two scales to hold beyond the one-loop level but we have no rigorous proof of this at the moment. All these functions of $\Q$, being associated with a phase transition, are not analytic at $\Q=0$. In $d=4$, non-analyticities appear starting at order $\Q^{5/2}$ in $f^2_\pi$, and at order $\Q^{3/2}$ in  $c^2_s$ and $\mu $, as a consequence of the non-analytic terms in the low-density limit of $P(X)$~\cite{Joyce:2022ydd}.

We can  also derive a universal scaling relation for $f_\pi$ at high density ($\Q\rightarrow \infty$) under the assumption that the ($\mu=0$) theory flows to a Conformal Field Theory (CFT) at high energies, and that the superfluid EFT at high densities is that of a conformal superfluid up to scaling violations (see also~\cite{Creminelli:2022onn}). This is a nontrivial assumption, being violated for instance in the case of super-renormalizable theories. Indeed, even if the $\mu=0$ theory flows towards a free theory in the UV, the chemical potential term always destabilizes the free theory ground state, giving a prominent role to the interaction terms. If only relevant couplings are present, as in a super-renormalizable theory, the superfluid EFT will not flow to the EFT of a conformal superfluid, but will display a different scaling behavior (see for instance the case of the ${\rm O}(N)$ model in $d=3$ of section~\ref{sec:ON}).

In the case of an almost conformal superfluid, the sound speed can be expressed as:
\begin{equation}\label{eq:conf_soundspeed}
c_s^2 (Q)= \dfrac{1}{d-1} + \Delta c_s^2(Q),
\end{equation}
where~\cite{Joyce:2022ydd}:
\begin{subequations}
\begin{align}
&\Delta c_s^2 = - \dfrac{1 }{(d-1)} \dfrac{T'(\mu^2) }{T'(\mu^2)+(d-1)P'(\mu^2)} \; ,\\
&T(\mu^2) \equiv T^\mu {}_\mu(\mu^2)=2 P'(\mu^2)\mu^2- d \, P(\mu^2),
\end{align}
\end{subequations}
so that at large densities (large $\mu$), scaling violations, being proportional to the beta function of the couplings, are small, and $\Delta c_s^2 \ll 1$. 
Neglecting scaling violations, solving eq.~\eqref{eq:dfpi} we get the universal behavior
\begin{equation}
f_\pi \sim \Q^{\frac{1}{d-1}} \qquad \mbox{for } \Q\rightarrow \infty \; ,
\end{equation}
which in fact just follows from dimensional analysis, since in a conformal superfluid $Q$ is the only independent dimensionful quantity. We can be more precise. Using \eqref{eq:conf_soundspeed} in \eqref{f(Q) general} we can write
\begin{equation}\label{eq:scaling_asymp}
f_{\pi}^{d-2}(\Q) = e^{- K_0} \, \dfrac{\Q_0^{\frac{1}{d-1}}}{m_{\rm pole}}  \, \Q^{\frac{d-2}{d-1}} \Big( 1 + \mathcal{O}(\epsilon)   \Big), \qquad \qquad \Q\rightarrow \infty
\end{equation}
where
\begin{equation}
K_0 = K(\Q_0)\equiv \int_0^{\Q_0} c_s^2 \,\dfrac{{\rm d}\Q}{\Q}  + \int_{\Q_0}^{\infty}\Delta c_s^2 \, \dfrac{{\rm d}\Q}{\Q} ,
\qquad  \epsilon \equiv \int_\Q^\infty \Delta c_s^2 \,\dfrac{{\rm d}\Q'}{\Q'} ,
\end{equation}
and $Q_0$ is an arbitrary reference charge density.
These quantities are always finite, thanks to the assumed asymptotic behavior of $c_s^2(Q)$. Despite  appearances, the above  expression for $f_\pi$ is independent of the choice of $\Q_0$, as  can be easily checked by taking a derivative with respect to $\Q_0$. 

One convenient choice of $\Q_0$ ($\bar Q$), can be defined by the condition $K(\bar Q)=0$, corresponding to $\bar \Q= e^{-(d-1)K(\Q_0)}\Q_0$, for any other $Q_0$,\footnote{This implies that the function $K(x)$ satisfies the functional equation $K(x \,e^{-(d-1)K(x)})=0$. Defining $h(x)= x \,e^{-(d-1)K(x)}$ this corresponds to $h(h(x))=h(x)$, that has $h(x)=\rm const$ or $h(x)=x$ as smooth solutions, as it can be readily proved by taking a derivative. The constant solution corresponds to the physical property we already noticed: eq.~\eqref{eq:scaling_asymp} is independent of $\Q_0$.} so that eq.~\eqref{eq:scaling_asymp} simply becomes
\begin{equation}
f_{\pi}^{d-2}(\Q) \simeq  \dfrac{ \big (\bar Q \Q^{d-2})^{\frac{1}{d-1}}}{m_{\rm pole}} \; , \qquad \qquad \Q\rightarrow \infty
\end{equation}

Another particularly natural choice ($ Q_\star$) is suggested by dimensional analysis. In $d$ dimensions, if there is only one marginal coupling $\lambda \phi^\alpha$, where $\alpha=2d/(d-2)$, and one mass scale (the pole mass $m_{\rm pole}$), keeping track of units of action $[S]$, one has
\begin{equation}
\label{eq:Qstar}
\begin{cases}
[\phi]= [m]^{(d-2)/2} [S]^{1/2},\\[10pt]
[\lambda]= [S]^{-2/(d-2)},
\end{cases}
\implies 
\qquad
\begin{cases}
f^{d-2} _\star = \dfrac{m_{\rm pole}}{\sqrt{\lambda}}, \\[10pt]
 \Q_\star = \dfrac{m_{\rm pole}^{d-1}}{\lambda^{(d-2)/2}}.
\end{cases}
\end{equation}
As we shall see in the explicit example of $\lambda \phi^4$ in $d=4$, these are exactly the values of $f_{\pi}$ and $\Q$ where the scaling behavior changes. Choosing $\Q_0=\Q_\star$ and plugging these values into eq.~\eqref{eq:scaling_asymp} we arrive at
\begin{equation}
\frac{f_{\pi}}{f_\star} \simeq e^{- \frac{1}{d-2}K_\star} \left(\dfrac{\Q}{Q_\star} \right)^{\frac{1}{d-1}}, \quad {\rm or \;\,  equivalently} \quad f_{\pi}\simeq e^{- \frac{1}{d-2}K_\star} \left(\dfrac{\Q}{\lambda} \right)^{\frac{1}{d-1}} , \qquad  \Q\rightarrow \infty,
\end{equation}
where $K_\star= K(\Q_\star)$. A redefinition of $\lambda$ is always compensated by a change of the value of $K_\star$ and a particularly physical choice can be made by defining the coupling in terms of a physical scattering amplitude at threshold~\cite{Joyce:2022ydd}.

On very general grounds, positivity and subluminality of $c_s^2$ can be used directly in \eqref{eq:dfpi} to derive strict bounds on the behavior of $f_\pi$ as a function of $\Q$.
The positivity of $c_s^2$ immediately implies the general upper bound 
\begin{equation}
f_\pi^{d-2}(\Q) \leq \dfrac{\Q}{m_{\rm pole}}.
\end{equation}
Subluminality, $c_s^2\leq1$, can  instead be used to derive a monotonicity bound, \begin{equation}
f_\pi (\Q_1) \geq f_\pi(\Q_2) \qquad \mbox{for } Q_1 > Q_2 \; .
\end{equation}
More in detail, $f_\pi(\Q)$ is a monotonically increasing function of $\Q$ with bounded derivative:
\begin{equation}
0 \leq \dfrac{{\rm d}\log f_\pi}{{\rm d}\log \Q} \leq \frac{1}{d-2} \; .
\end{equation}
In particular, $({\rm d}f_\pi^{d-2} / {\rm d}\Q) \leq 1/m_{\rm pole}$ is everywhere satisfied and is a strict bound everywhere except at $\Q=0$, where it is saturated.

\section{The ${\rm O}(N)$ model at large 
$N$}
\label{sec:ON}

We have seen that the relation between finite density and symmetry breaking holds quite generally at one loop for $\rm U(1)$ symmetric theories of a complex scalar field. 
We wish now to give additional support for the conjectural relationship between these two phenomena by proving that it holds non-perturbatively in the large $N$ limit for a ${\rm O}(N)$ scalar theory. The model is that of $N$ real scalar fields transforming in the vector representation and interacting with quartic interactions in $d=3$.\footnote{This theory and its quantum effective potential for $\mu=0$ have been analyzed at large $N$ in~\cite{Coleman:1974jh}. The related ${\rm O}(N)$ model with quartic interactions in four dimensions is known to be plagued by a tachyonic instability~\cite{Gross:1974jv,Coleman:1974jh}, see also~\cite{Abbott:1975bn}. For a textbook treatment see \emph{e.g.}~\cite{Zinn-Justin:1989rgp}.} Related discussions on ${\rm O}(N)$ vector models at finite $\mu$ and their relation to the large charge expansion have appeared, for instance, in Refs.~\cite{Alvarez-Gaume:2019biu,Orlando:2021usz,Giombi:2022gjj}.

The finite $\mu$ Lagrangian can be computed along the lines previously described. To be concrete let us denote by $\vec{\Sigma}$ a set of $N$ scalar fields transforming in the vector (fundamental) representation of ${\rm O}(N)$, and assume that we have chosen a basis such that we have a chemical potential for the charge associated with rotations in the $(\Sigma_1,\Sigma_2)$ plane. It is convenient to adopt the following notation: $\Sigma_I = (s_1,s_2, S_i)$, where $I=1,\dots, N$ and $i=3,\dots,N$. In this notation we have:
\begin{equation}\label{eq:lagrangian_ON}
\begin{split}
\mathcal L_{\mu} = &\, \dfrac{1}{2} (\partial_{\nu}s_{1})^{2} +\dfrac{1}{2} (\partial_{\nu}s_{2})^{2} +\mu \left( \dot{s_{1}}s_{2}- \dot{s_{2}}s_{1}\right) - \dfrac{1}{2} (m^2-\mu^2) (s_{1}^2+s_{2}^2) - \dfrac{\lambda_N}{4N}(s_{1}^2+s_{2}^2)^2\\
&+ \dfrac{1}{2} (\partial_{\nu}S_{i})^{2} - \dfrac{1}{2} \left[ m^2+\lambda  (s_{1}^2+s_{2}^2) \right] S_{i}^{2} - \dfrac{\lambda_N}{4N} S_{i}^{4},
\end{split}
\end{equation}
where we introduced $\lambda_N = \lambda N$. We work in the limit of large $N$ and small $\lambda$, with $\lambda_N$ fixed, but to all orders in $\lambda_N$. Moreover, for simplicity let us assume $m^2\geq0$ (the  $m^2 < 0$ case can be treated similarly). This model can be solved at large $N$ introducing an auxiliary field $\chi$~\cite{Gross:1974jv,Coleman:1974jh}.
To do this we add to the Lagrangian the  Gaussian term
\begin{equation}
\delta\mathcal{L} =\dfrac{N}{4\lambda_N} \left( \chi - m^2 - \dfrac{\lambda_N}{N} \Sigma_I^2\right)^2 ,
\end{equation}
which has the only effect of changing the normalization of the path integral by a constant.
Carrying out the algebra we see that the quartic interaction terms are canceled by the new auxiliary term, and that the only residual interactions are trilinear couplings with the auxiliary field $\chi$:
\begin{equation}\label{eq:lagrangian_ON_aux}
\begin{split} 
\mathcal L_{\mu} =&\, \dfrac{1}{2} (\partial_{\nu}s_{1})^{2} +\dfrac{1}{2} (\partial_{\nu}s_{2})^{2} +\mu \left( \dot{s_{1}}s_{2}- \dot{s_{2}}s_{1}\right) - \dfrac{1}{2} (\chi -\mu^2) (s_{1}^2+s_{2}^2) \\
&+ \dfrac{1}{2} (\partial_{\nu}S_{i})^{2} - \dfrac{1}{2} \chi S_{i}^{2} + \dfrac{N}{4\lambda_N} \left( \chi - m^2 \right)^2 .
\end{split}
\end{equation}
We are interested in computing the effective potential as a function of $s_{1,2}$, $S_i$ and $\chi$ in the large $N$ limit.
From the new Lagrangian, we see that the $\chi$ propagator is suppressed by a factor $1/N$. As a consequence, it is easy to see from diagrammatic arguments that the only quantum contributions to the effective potential that are of the same order as the tree level terms, in the $1/N$ expansion,  arise from one-loop graphs with $s_{1,2}$ and $S_i$ internal propagators, and no internal $\chi$.\footnote{The first line in eq.~\eqref{eq:lagrangian_ON_aux} is formally of order $1$, while the second line is of order $N$ (with $\chi$ of order $1$). We are interested in the leading order contributions in the effective potential for both $s_{1,2}$ and $S_i$, and we neglect consistently subleading orders in $1/N$ generated when including loop diagrams with virtual $\chi$'s. In fact, in the symmetry broken phase it will turn out that $s_i^2 \sim N$.} These contributions can be computed exactly, since the action written in terms of the auxiliary field is quadratic in $s_{1,2}$ and $S_i$. The full result at leading order in $1/N$ in the $\overline{\text{MS}}$ scheme is
\begin{equation}\label{eq:eff_potential_SON}
V_{\rm eff}(s,S,\chi;\mu) =\dfrac{1}{2} (\chi -\mu^2) (s_{1}^2+s_{2}^2) + \dfrac{1}{2} \chi S_{i}^{2} - \dfrac{N}{4\lambda_N} \left( \chi - m^2 \right)^2 - \dfrac{N}{12\pi} \chi^{3/2} .
\end{equation}

The finite $\mu$ ground state of the theory corresponds to a stationary point of $V_{\rm eff}(s,S,\chi;\mu)$, which is determined by the conditions
\begin{subequations}\label{eq:conditions}
\begin{align}
& \left(s_{1}^2+s_{2}^2  \right)  + S_{i}^{2} = \dfrac{N}{\lambda_N} \left( \chi - m^2 \right) +  \dfrac{N}{4\pi} \sqrt{\chi}, \label{eq:condition1}\\
& \left( \chi - \mu^2 \right) s_{1,2} = 0, \\
&\;\; \chi\,  S_{i} = 0 .
\end{align}
\end{subequations}
From eq.~\eqref{eq:condition1} it follows the in the ${\rm O}(N)$ symmetric state with $s_{1,2}=S_i=0$, $\chi$ is a function of $(m,\lambda_N,N)$ and is independent of $\mu$. As a consequence, the value of the potential~\eqref{eq:eff_potential_SON} for such a state is independent of $\mu$ and cannot support finite density. Therefore finite density is always accompanied by spontaneous symmetry breaking. More in detail, as we discuss in Appendix~\ref{app:SON}, when $\mu^2 < \mu^2_{\rm crit}$ the potential is minimized for $s_{1,2}=S_i=0$, the ${\rm O}(N)$ symmetry is unbroken and the charge density is zero. On the other hand, whenever $\mu > \mu_{\rm crit}$ the minimum of the effective potential is attained for\footnote{The global stability of this minimum is valid in $d=3$, as discussed in Appendix~\ref{app:SON}. The similar result in $d=4$ is plagued by an instability at large values of $\chi$.}
\begin{subequations}
\begin{align}
&\left(s_{1}^2+s_{2}^2  \right) =  \dfrac{N}{\lambda_N} \left( \mu^2 - m^2 \right) +  \dfrac{N}{4\pi} \mu , \label{eq:solution1}\\
& \chi = \mu^2 , \\
&S_i=0 .
\end{align}
\end{subequations}
The critical value of $\mu$ can be found from eq.~\eqref{eq:solution1} and the condition $\left(s_{1}^2+s_{2}^2  \right)\geq 0$. We find
\begin{equation}
\label{eq:mucrit_largeN}
\mu^2_{\rm crit} =\left( \sqrt{\dfrac{\lambda_N^2}{64\pi^2}+m^2} - \dfrac{\lambda_N}{8\pi}\right)^2,
\end{equation}
which coincides with the pole mass $m^2_{\rm pole}$ in the unbroken phase of the UV theory with $\mu=0$, as shown in Appendix~\ref{app:SON}. Notice that this result is non-perturbative in $\lambda_N$.

For $\mu > \mu_{\rm crit}$, the value of the effective potential at its minimum is
\begin{equation}
V_{\rm min}(\mu) = - \dfrac{N}{4\lambda_N} \left( \mu^2 - m^2 \right)^2 - \dfrac{N}{12\pi} \mu^{3},
\end{equation}
so that from eq.~\eqref{eq:chargevev} the charge density is 
\begin{equation}\label{eq:Jmu}
Q = \dfrac{N}{\lambda_N} \mu (\mu^2-m^2) +  \dfrac{N}{4\pi} \mu^2 .
\end{equation}
In particular the following relation is satisfied: $\Q= \mu \left(s_{1}^2+s_{2}^2  \right)$, where $s_{1,2}$ denote the expectation values of the corresponding quantum operators on the ground state.

From the same result we obtain the low-energy effective action for the superfluid phase of this model at leading order in $N$:
\begin{equation}\label{eq:PX_largeN}
P(X)\Big\vert_{N\rightarrow \infty} = \dfrac{N}{4\lambda_N} \left(X - m^2 \right)^2 + \dfrac{N}{12\pi} X^{3/2} + \mathcal{O}(1) \qquad \qquad(d=3).
\end{equation}
We can compare this result with that obtained by repeating the computation of~\cite{Joyce:2022ydd} in $d=3$, working at finite $N$ but only at one loop in $\lambda_N$ (or equivalently $\lambda$):
\begin{equation}
\begin{split}
P(X)\Big\vert_{\rm one-loop}= & \, \dfrac{N}{4\lambda_N} \left(X - m^2 \right)^2 \\
&+ \dfrac{1}{48\pi} \left[ \left(9X -5 m^2\right) \sqrt{6X - 2m^2 } + 3 \dfrac{\left(X - m^2 \right)^2}{\sqrt{X}} \,{\rm arcsinh}\left(\sqrt{\dfrac{2X}{X - m^2 }}\right) \right] \\ 
&+ \left(\dfrac{N-2}{12\pi}\right) X^{3/2} ,
\end{split}
\end{equation}
where the first line is the tree-level contribution, the second line is the one-loop correction coming from integrating out the radial mode and the third line is that arising from integrating out the $(N-2)$ gapped Goldstones.
We see that at large $N$ the effective Lagrangian coincides with that of eq.~\eqref{eq:PX_largeN}: at leading order in $1/N$ the large $N$ result is one-loop exact.

The critical value of $\mu$ \eqref{eq:mucrit_largeN} matches that  inferred from a consistency analysis of the low-energy $P(X)$ theory. Requiring stability and subluminality of the phonon perturbations~\cite{Joyce:2022ydd} gives the conditions
\begin{equation}
P'(\mu^2)>0 \quad {\rm and} \quad P''(\mu^2)>0,
\end{equation}
which are satisfied exactly for $\mu>\mu_{\rm crit}$.

We can now analyze this result in light of the scaling relations of section~\ref{sec:scaling_relations}. From~\eqref{eq:PX_largeN} we have 
\begin{equation}
f_\pi = 2 P'(X)= \dfrac{N}{\lambda_N} \left( \mu^2 - m^2 \right) +  \dfrac{N}{4\pi} \mu,
\end{equation}
with $\Q$ and $\mu$ related by eq.~\eqref{eq:Jmu}. The small $\Q$ limit gives $f_\pi\simeq \Q/m_{\rm pole}$ as expected. The high density limit of this model is peculiar in that the theory we are studying is super-renormalizable in $d=3$. As a consequence, the superfluid EFT we obtain does not approach the EFT of a conformal superfluid even if the $\mu=0$ theory is conformal in the UV. 
More explicitly, we find 
\begin{equation}
P(X)\Big\vert_{N\rightarrow \infty} \simeq \dfrac{N}{4\lambda_N} X^2, \quad f_\pi \simeq \dfrac{N^{1/3}}{\lambda_N^{1/3}}\Q^{2/3},  \qquad \Q\rightarrow \infty,
\end{equation}
where the coupling $\lambda_N$ has mass dimension one, so that the $P(X)$ and the asymptotic scaling are not those of a $d=3$ conformal superfluid.

We can formally carry out the same analysis also in $d=4$, by considering the symmetry breaking minimum of the effective potential. This is only a local minimum, as the true ground state is the symmetric one~\cite{Abbott:1975bn}. However the instability is non-perturbative, so that the symmetry breaking ground state is metastable, at least in the limit of small $\lambda_N$. Repeating the previous analysis, we find:
\begin{equation}\label{eq:eff_potential_SON_d4}
V_{\rm eff}(s,S,\chi;\mu) =\dfrac{1}{2} (\chi -\mu^2) (s_{1}^2+s_{2}^2) + \dfrac{1}{2} \chi S_{i}^{2} - \dfrac{N}{4\lambda_N} \left( \chi - m^2 \right)^2 - \dfrac{N}{384\pi^2} \chi^{2} (9-6\log(\chi)),
\end{equation}
\begin{equation}
V_{\rm min}(\mu) = - \dfrac{N}{4\lambda_N} \left( \mu^2 - m^2 \right)^2 - \dfrac{N}{384\pi^2} \mu^{4} (9-6\log(\mu^2)),
\end{equation}
so that the superfluid EFT at large $N$ is
\begin{equation}\label{eq:PX_largeN_d4}
P(X)\Big\vert_{N\rightarrow \infty} = \dfrac{N}{4\lambda_N} \left(X - m^2 \right)^2 + \dfrac{N}{384\pi^2} X^{2} (9-6\log(X)) + \mathcal{O}(1) \qquad (d=4).
\end{equation}
At high density we can resum the logs using the running coupling, along the lines of~\cite{Joyce:2022ydd}. The result is
\begin{equation}\label{eq:PX_largeN_d4}
P(X)\Big\vert_{N\rightarrow \infty} \simeq \dfrac{N}{4\lambda_N(X)} X^2 \qquad \Q\rightarrow \infty \qquad\qquad (d=4).
\end{equation}
The scaling of $f_\pi$ thus is
\begin{equation}
f_\pi^2 \simeq \dfrac{N^{1/3}}{\lambda_N^{1/3}}\Q^{2/3}.
\end{equation}
This suggests that in the (conformal) superfluid phase of a $d=4$ CFT at large $N$ (see \emph{e.g.}~\cite{Alvarez-Gaume:2019biu,Orlando:2021usz} for some examples in $d=3$) the symmetry breaking scale $f_\pi$ scales as $f_\pi^2 \sim N^{1/3} $, where $N$ is a parameter counting the number of species charged under the ${\rm U}(1)$ symmetry associated to $\mu$. Our calculation is valid to all orders in $\lambda_N$ (at leading order in $1/N$), suggesting that this scaling could be valid also in strongly coupled CFTs. It would be interesting to explore the regime of validity of such a relation in the framework of Conformal Field Theory (and its generalization to other dimensions), and understand how $N$ is encoded in the CFT data (for instance if it is related to a combination of OPE coefficients in the current-current OPE).\footnote{We thank Gabriel Cuomo for discussions on this point.}

\section{Applications and relation to previous works}
\label{sec:app}

\subsection{Conformal superfluid in $d=3$: massless $\lambda \phi^6$}

In order to make contact with existing results and confirm the validity of our computations we can consider some special cases. 
Let us start from the three-dimensional theory of a massless complex scalar with $\lambda \phi^6$ interactions. At the classical level this theory is scale (and conformal) invariant, since the $\lambda \phi^6$ interaction is marginal and the field is massless. This property continues to hold also at one loop, for any perturbative value of $\lambda$ and in $d=3$, as pointed out in~\cite{Badel:2019khk}, since the running of $\lambda$ first arises  at two loops. This theory has also been considered as an interesting benchmark model in relation to positivity bounds in theories with non-linearly realized Lorentz invariance~\cite{Creminelli:2022onn}.
In our approach, the absence of running at one loop is a consequence of the absence of logarithmic divergencies, see eqs.~\eqref{eq:Veff_g_d} and~\eqref{eq:Veff_mu_d}. More generally, we see that this holds for arbitrary potentials in odd dimensions.

We therefore consider the $\lambda \phi^6$ theory at finite $\mu$ in $d=3$, with classical potential
\begin{equation}
V(\phi;\mu)= -\mu^2 \phi^2 + \lambda \phi^6 ,
\end{equation}
for which $g_{\rm min}=2\mu^2$ and $\left(\phi_{0}\right)^4=\mu^2/(3\lambda)$.
From eq.~\eqref{eq:PX_3d} we immediately obtain
\begin{equation}
P(X)\Big\vert_{d=3,\lambda\phi^6}= \left( \dfrac{2}{3^{3/2} \, \sqrt{\lambda}} + \dfrac{7\sqrt{2}+3 \,{\rm arcsinh}(1)}{12\pi}\right) X^{3/2},
\end{equation}
which has the form expected for a conformal superfluid.\footnote{Notice that ${\rm arcsinh}(1)$ can be equivalently expressed as $\log(1 + \sqrt{2})$.}
We can compare this result with that of~\cite{Badel:2019khk} by taking into account the different conventions for the coupling constant.\footnote{There is a minus sign misprint in eq.~(34) of Ref.~\cite{Badel:2019khk}. We thank A. Monin for double checking.} Denoting their coupling as $\hat{\lambda}$, the relation is $\lambda=\hat{\lambda}^2/36$, and we find that in terms of their notation
\begin{equation}
\alpha_1 = \dfrac{4}{\sqrt{3}} + \dfrac{7\sqrt{2}+3 \,{\rm arcsinh}(1)}{12\pi} \hat{\lambda} =  \dfrac{4}{\sqrt{3}} + 0.33273\dots  \times \hat{\lambda},
\end{equation}
to be compared with $\alpha_1 = 4/\sqrt{3} + 0.3326 \; \hat{\lambda}$, where $P(X)= \alpha_1 X^{3/2}/\hat{\lambda}$.\footnote{We neglect corrections to $\alpha_1$ of order $\mathcal{O}(\hat{\lambda}^2)$ or higher.}

As discussed in~\cite{Hellerman:2015nra,Monin:2016jmo,Badel:2019oxl} (see~\cite{Gaume:2020bmp} for a review), from the one-loop effective action for the superfluid phase of $\lambda \phi^6$ it is possible to extract the scaling dimension of the lowest dimensional charge-$n$ operator (denoted as $\Delta_{\phi^n}$), in the limit of large charge $n$. In the notation of~\cite{Badel:2019khk}:
\begin{equation}
\Delta_{\phi^n} = \left(\dfrac{\hat{\lambda} n}{\sqrt{3}\pi}\right)^{3/2} \left[ c_{3/2} + \mathcal{O}\left(\dfrac{\sqrt{3}\pi}{\hat{\lambda} n}\right)\right],
\end{equation} 
and $\hat{\lambda} \, c_{3/2} = \pi/ (3^{3/4} \sqrt{\alpha_1})$. We find
\begin{equation}
c_{3/2} = \dfrac{\sqrt{3}\pi}{6\hat{\lambda}} - \dfrac{7\sqrt{2}+3 \,{\rm arcsinh}(1)}{192} =  \dfrac{\sqrt{3}\pi}{6\hat{\lambda}} - 0.0653313 \dots  ,
\end{equation} 
in perfect numerical agreement with the result of~\cite{Badel:2019khk} within its accuracy.

\subsection{Quartic potential in $d=4$}

In section~\ref{sec:gexpansion} we computed the effective potential for a complex scalar field with arbitrary interaction potential $V_{\rm int}(\phi)$, in generic $d>2$ dimension, by expanding in powers of $g$ defined in eq.~\eqref{eq:M2g2_general}. Here, we want to go back to our result \eqref{eq:Veff_g_4} and specialize to the particular case 
\begin{equation}
V_{\rm int}(\phi)=\lambda \phi^4, \qquad\qquad (d=4).
\end{equation}
In this case one has $M^2 = m^2-\mu^2 +4 \lambda \phi^2$ and $g^2 = 4 \lambda^2 \phi^4$.
It is straightforward to check that the counterterms needed to cancel UV divergencies~\eqref{eq:poles} are in perfect agreement with the standard dimensional regularization results for the $\phi^{4}$ theory with $\mu=0$.
Working in the $\overline{\text{MS}}$ scheme (with $\bar\mu=1$ for simplicity) we obtain the result:
\begin{equation}
\begin{split}
V_{\rm eff}^{(1)}(\phi;\mu)\Big\vert_{d=4}  = &-\dfrac{3}{64\pi^{2}} \left( m^{2}+4\lambda\phi^2 \right)^{2} \left(1-\dfrac{2}{3}\log\left(m^{2}+4\lambda\phi^2 \right)\right) \\
&- \dfrac{\lambda^{2}}{4\pi^{2}} \phi^4 \left(1-\dfrac{1}{2}\log\left(m^{2}+4\lambda\phi^2\right) -\dfrac{M}{\mu}\arctan\left(\dfrac{\mu}{M}\right)\right) \\
&- \dfrac{\lambda^{2}}{64\pi^{2}} \phi^4 \int_{0}^{1}{\rm d}x \,y \;{}_3F_2\left[\begin{array}{c}
1,~~1,~~\frac{3}{2}\\
2,~~3
\end{array}\Bigg\vert y \right]
\; ,
\end{split}
\end{equation}
where now
\begin{equation} 
y= \dfrac{16 x (1-x) \lambda^{2} \phi^4}{(M^{2}+(1-2x)^{2}\mu^{2})^{2}}.
\end{equation}

Expanding for $\lambda \phi^2 \ll m^{2} $, assuming $\mu \neq m$, and adding the tree-level contribution, we obtain the effective potential at one loop:
\begin{equation}
\begin{split}
V_{\rm eff}(\phi)=& -\dfrac{3 m^{4}}{64\pi^{2}} \left(1-\dfrac{2}{3}\log m^2 \right)+ \left(m^{2}-\mu^{2} - \dfrac{\lambda m^{2}}{4\pi^{2}} \left(1- \log m^2\right)\right) \phi^2 + \\
&\hspace{-20pt}+ \left(1 - \dfrac{\lambda}{8\pi^{2}} \left(2 - 5\log m^2 - 2 \dfrac{\sqrt{m^{2}-\mu^{2}}}{\mu}
\arctan\left( \dfrac{\mu}{\sqrt{m^{2}-\mu^{2}}}\right) \right)\right) \lambda \phi^4 +
\mathcal{O}(\phi^{6}).
\end{split}
\end{equation}
Some comments are in order: first, notice that at $\phi=0$ the effective potential is $\mu$-independent, in agreement with the general results of section~\ref{sec:ssb}; second, the quadratic term identifies the critical value of $\mu$ with the pole mass computed in the $\overline{\rm MS}$ scheme as expected from the argument of section~\ref{sec:criticalmu}.

The $P(X)$ one-loop effective theory for the superfluid phonons can be obtained from eq.~\eqref{eq:PX_4d}, where now $g_{\rm min} = (X-m^2)$:
\begin{equation}
\begin{split}
P(X)\Big\vert_{\phi^4}= & \dfrac{(X-m^2)^2}{4\lambda} \\
&+ \dfrac{1}{192\pi^2} \Bigg( 
- 4 X^2 + \left(X^2 +2 g_{\rm min} X + 2 g_{\rm min}^2 \right) \left(9 - 6\log(2 g_{\rm min})\right)\\
&
\hspace{50pt} -\dfrac{5X^3}{2g_{\rm min}} \;{}_3F_2\left[\begin{array}{c}
1,~~1,~~\frac{7}{2}\\
4,~~5
\end{array}\Bigg\vert -\dfrac{2X}{g_{\rm min}} \right]
\; 
 \Bigg).\\ 
\end{split}
\end{equation}
and matches exactly the independent computation of Ref.~\cite{Joyce:2022ydd}, as we already noticed in section~\ref{sec:super_eft}.
We can also consider the low-density limit, corresponding to $X \simeq m_{\rm pole}^2$, from which we obtain (setting the renormalization $\bar{\mu}=m_{\rm pole}$):
\begin{equation}
P(X) =  \dfrac{(X-m_{\rm pole}^2)^2}{4\hat\lambda} - \dfrac{\sqrt{2}}{15\pi^2} \dfrac{(X-m_{\rm pole}^2)^{5/2}}{m_{\rm pole}} - \dfrac{1}{12\pi^2} \dfrac{(X-m_{\rm pole}^2)^{3}}{m_{\rm pole}^2} + \dots,
\label{PlowQ}
\end{equation}
where $\hat\lambda$ is defined in terms of the coupling at threshold~\cite{Joyce:2022ydd}
\begin{equation}
\dfrac{1}{\hat\lambda} = \dfrac{1}{\lambda_{\rm thr}} - \dfrac{1}{6\pi^2}, \qquad \qquad \lambda_{\rm thr}= \lambda_{\rm \overline{MS}} - \dfrac{5}{4\pi^2} \lambda_{\rm \overline{MS}}^2 \left(\dfrac{1}{3} - \log m\right), 
\end{equation}
where $\lambda_{\rm thr}$ is the two-to-two elastic scattering amplitude for identical particles at threshold in the $\mu=0$ theory:\footnote{This quantity can be easily rewritten in terms of the scattering length often used in the study of low-energy quantum mechanical scattering.}
\begin{equation}
i \mathcal{M}_{2\rightarrow 2}(s=4m_{\rm pole}^2, t=0, u=0) \equiv -6i \, \lambda_{\rm thr}.
\end{equation}

It can be instructive to study the scaling relations of section~\ref{sec:scaling_relations} in this explicit example. We notice that $P''(m^2_{\rm pole})=  1/2 \hat\lambda $, so that at low densities we have
\begin{subequations}
\begin{align}
& c_s^2(\Q)= \dfrac{\hat\lambda}{2m_{\rm pole}^3} \Q+ \dfrac{\hat\lambda^{5/2}}{2^{3/2}\pi^2 m_{\rm pole}^{9/2}} \Q^{3/2} + \dots, \\
& f_\pi^2(\Q)= \dfrac{\Q}{m_{\rm pole}} - \dfrac{\hat\lambda}{2 m_{\rm pole}^4} \Q^2 - \dfrac{\hat\lambda^{5/2}}{3  \sqrt{2}\pi^2 m_{\rm pole}^{11/2}} \Q^{5/2} + \dots, \\
& \mu(\Q) = m_{\rm pole} + \dfrac{\hat\lambda}{2m_{\rm pole}^2 } \Q+ \dfrac{\hat\lambda^{5/2}}{3  \sqrt{2}\pi^2 m_{\rm pole}^{7/2}} \Q^{3/2} + \dots,
\label{muQ}
\end{align}
\end{subequations}
where we only reported the leading order non-analyticities arising at one loop from the $(X-m_{\rm pole}^2)^{5/2}$ term. The subleading results can be computed straightforwardly (see appendix~\ref{app:details}).
On the other hand, at high densities we obtain the expected scaling for a superfluid theory which is approximately that of a conformal superfluid. Eq.~\eqref{eq:scaling_asymp} simplifies for the choice of $Q_{\star}$ defined in eq.~\eqref{eq:Qstar}, since $K_{\star}=0$ at tree level, and the high density scaling takes the simple form
\begin{equation}
f_{\pi}(\Q) \simeq \left(\dfrac{\Q}{\sqrt{\lambda}} \right)^{\frac{1}{3}}, \qquad \qquad \Q\rightarrow \infty.
\end{equation}

\subsection{The Lee--Huang--Yang relation and its relativistic extension}

As a last application, we shall compute the energy density of our $d=4$, $\phi^4$ theory in the limit of low charge density, which, for a massive theory,  is a non-relativistic limit. In fact, in the non-relativistic limit dilute systems of interacting bosons have been studied extensively. Lee, Huang and Yang~\cite{Lee:1957zza,Lee:1957zzb} showed that at low densities the energy density is organized as an expansion in powers of $\sqrt{Q a^3}$, where $Q$ is the charge density and $a$ is the scattering length, and computed the first correction to the leading order result for a dilute gas of hard spheres.\footnote{For a modern perspective and a discussion of higher order corrections see \emph{e.g.}~\cite{Braaten:1996rq,Braaten:1997ky} and the review~\cite{Andersen:2003qj}.} More generally, the Lee--Huang--Yang relation gives a rigorous lower bound on the energy density of dilute systems of non-relativistic interacting bosons, provided that the potential satisfies some  mild regularity conditions~\cite{fournais2020energy,fournais2022energy}. We shall reproduce here the Lee--Huang--Yang relation and compute relativistic corrections to it.

The energy density can be obtained from the time component of the energy-momentum tensor:
\begin{equation}
\rho= T^{00} = 2 P'(X)  X - P(X) = \Q \mu - P(\mu^2).
\end{equation}
Before discussing the explicit result, let us make some general comments based on dimensional analysis that can help in understanding the systematics of the low density expansion. We reintroduce factors of the speed of light $c$, and look for dimensionless combinations of the three quantities at our disposal ($\hat\lambda$, $m_{\rm pole}$, $Q$). It is convenient to trade $\hat\lambda$ for a length scale $\hat a$ (related to the scattering length by an order one number), defining $\hat\lambda = \hat a \,m_{\rm pole} \,c$. There is only one dimensionless combination that is independent of the speed of light $c$, and another one can be taken to parametrize relativistic corrections. By studying the structure of the loop expansion one can identify the two expansion parameters as
\begin{equation}
\kappa=\sqrt{Q \hat a^3} ,\quad \xi = \dfrac{\sqrt{Q \hat a}}{m_{\rm pole}\, c},
\end{equation}
where $\kappa$ acts as a loop counting parameter and $\xi$ parametrizes relativistic corrections.\footnote{In order to arrive at this result notice that the relativistic expansion parameter $\xi^2$ can be identified as the leading order non-relativistic chemical potential $\mu_{\rm NR} =\hat{a} Q/2m_{\rm pole}$ normalized by the rest energy $\xi^2=2\mu_{\rm NR}/m_{\rm pole}c^2$. Then $\kappa = \hat{\lambda} \xi$, and its origin as loop counting parameter is manifest.}
From \eqref{PlowQ} and \eqref{muQ}, we obtain 
\begin{equation}
\rho = Q m_{\rm pole}c^2 +  \dfrac{\hat{a} Q^2}{4 m_{\rm pole}} \left(1-\dfrac{1}{2}\xi^2 +\left(\dfrac{4\sqrt{2}}{15 \pi^2}+\dfrac{1}{3 \pi^2} \xi \right) \kappa + \mathcal{O}(\kappa^2)\right),
\end{equation}
where the first term represents the rest energy contribution, while the $\xi$ terms are relativistic corrections. Discarding the rest energy and setting $\xi=0$ we recover the result of Lee, Huang and Yang~\cite{Lee:1957zza,Lee:1957zzb} upon mapping $\hat{a} \rightarrow 8\pi a$, where $a$ is the rigid sphere scattering length:
\begin{equation}
\rho_{\rm NR} =   \dfrac{2\pi a Q^2}{m_{\rm pole}} \left(1 + \dfrac{128}{15} \sqrt{\dfrac{Q a^3}{\pi}} + \mathcal{O}\left(Q a^3\right)\right) \; .
\end{equation}

The coefficient of the subleading term is a non-trivial consistency check of our formalism, signaling that in the non-relativistic limit the $\lambda \phi^4$ model admits an effective description in terms of rigid spheres. This fact suggests that the low density physics has some degree of universality --- see eq.~\eqref{nonanalyticities}. We hope to come back to this issue in the future. We are not aware of previous computations of relativistic corrections to the Lee--Huang--Yang relation. It would be interesting to understand their general status, and in particular if they provide a lower bound on the relativistic energy density along the lines of what has been proved in the non-relativistic case~\cite{fournais2020energy,fournais2022energy}.

\section{Outlook: towards a non-perturbative understanding}
For ${\rm U}(1)$-symmetric scalar field theories in generic spacetime dimensions $d>2$, we have given strong evidence for these two statements:
\begin{enumerate}
\item
The ground state at finite chemical potential cannot develop a charge density unless it (spontaneously) breaks the ${\rm U}(1)$ symmetry;
\item
Assuming there is no SSB at zero chemical potential, the system develops charge density and SSB only when the chemical potential exceeds the pole mass of the charged particles in the Poincar\'e invariant vacuum.
\end{enumerate}
We have proved these facts in perturbation theory at one loop for generic non-derivative interactions, and non-perturbatively in the  ${\rm O}(N)$ model  with quartic interactions to leading order in $1/N$ (in $d=3$ and $4$).
It is interesting to consider how to go beyond these limits. Can we prove these facts in general, non-perturbatively?

If {\it i)} the theory's lightest states are spinless charged particles of nonzero mass $m_{\rm pole}$, and {\it ii)} these interact only through short range interactions, then a possible argument for very small charge densities goes as follows: 
if the density is so low that the average distance between the particles is much longer than the interactions' range and the particles' Compton wavelength, then one can assume that the particles are free; to leading order in these approximations, the results of a free scalar QFT should then apply. In particular, a nonzero density cannot arise for $\mu < m_{\rm pole}$. For $\mu = m_{\rm pole}$, Bose--Einstein condensation kicks in, leading to both a nonzero density and the spontaneous breaking of the ${\rm U}(1)$ symmetry \cite{Strocchi:2008gsa}. 

There is probably some truth in this argument, but it is not completely convincing to us, because, as we reviewed, the chemical potential is not a good control parameter in the free-theory case, since nothing happens for $\mu<m$, and there is no ground state for $\mu > m_{\rm pole}$. So, from this viewpoint the free-theory case is a degenerate limit and interactions, however weak or short range they might be, are important to stabilize the system for $\mu > m_{\rm pole}$. However, this might also suggest that working at fixed chemical potential rather than fixed density is a bad choice for certain questions.

A different approach is a functional one: Consider the finite-$\mu$ path-integral representation for the generating functional $W[J; \mu]$, where $J(x)$ is the source for $\Phi$. It is easy to see that, upon changing the integration variable as $\Phi = \Phi' e^{i \mu t}$, one can move the chemical potential to the source,
\be
W[J; \mu] = W[J e^{i \mu t}; 0] \; ,
\ee
as can also be understood by noticing that introducing a chemical potential is equivalent to modifying the generator of time translation as $H \to H - \mu Q$.
Now, this ties the charge density---the derivative of $W$ with respect to $\mu$---to the expectation value of $\Phi$---the derivative of $W$ with respect to $J$: the former cannot be nonzero if the latter vanishes.
This argument, however, must be too simplistic, because it would work essentially unaltered in the free Fermi gas case, where we know the ground state, and we know that there is no SSB at finite density. Indeed, for a free fermionic path integral at finite $\mu$, consider introducing a source $J(x)$ for a scalar charged operator, such as the Majorana mass combination $\psi \psi \equiv \psi^T \cdot i\gamma^2 \cdot \psi $, which has charge two. Then, the same manipulations on $W[J]$ as above would lead to the same conclusion: by redefining the integration variable as $\psi = \psi' e^{i \mu t}$, one can move the chemical potential to the source, $W[J; \mu] = W[J e^{2 i \mu t}; 0] $, showing that there cannot be a nonzero charge density if the expectation value of $\psi \psi$ vanishes. This conclusion conflicts with reality for the free Fermi gas, which suggests that this functional argument must be neglecting some important technical details.

Finally, one could consider the connection of our phenomenon with the spontaneous breaking of boosts. The  Goldstone theorem associated with these exhibits important differences with more standard ones, and can be obeyed in unconventional ways  \cite{Matsumoto:1986xs, Alberte:2020eil, Komargodski:2021zzy}. Perhaps one can show that, under reasonable assumptions, if the ground state of a relativistic system breaks boosts but no rotations or spatial translations, the gapless excitations required by the Goldstone theorem can only be of two types: the particle-hole continuum of a Fermi liquid, or the phonons of a superfluid. This would prove our conjecture, because a finite density certainly breaks boosts, and so in a homogeneous and isotropic bosonic system this would imply a superfluid-like low-energy spectrum.

We hope to make progress in these directions in the near future.

\acknowledgments

It is a pleasure to thank Paolo Creminelli, Gabriel Cuomo, Luca Delacr\'etaz, Lorenzo Di Pietro, Paolo Glorioso, Hofie Hannesdottir, Oliver Janssen, Austin Joyce, and Riccardo Rattazzi for useful discussions. We are especially grateful to Austin Joyce for early collaboration, to Paolo Creminelli for prompting the analysis of the $\lambda \phi^6$ model in $d=3$, and to Lorenzo Di Pietro for a question that inspired our study of the ${\rm O}(N)$ model. We also thank Jens Andersen, Gabriel Cuomo, Luca Delacr\'etaz and Sean Hartnoll for comments on a preliminary version of this work. AP is grateful to the audiences of the Sapienza University of Rome, ICTP, APC Paris, Saclay and Perimeter seminars, where part of this work was presented, for interesting comments and questions. The work of AP is supported by the grant DOE DE-SC0011941. LS is supported by the Centre National de la Recherche Scientifique (CNRS).

\appendix 

\section{The spinning rigid rotor and its ground state energy}
\label{app:rotor}

In this appendix we generalize the one-loop quantum mechanical effective potential to the case of a particle moving in three dimensional space, and use this result to rederive the well-known quantization condition for the ground state energy of a quantum mechanical rigid rotor at fixed angular momentum $\hat{L}_{3}=J$:
\begin{equation}
E_{0}= \dfrac{J(J+1)}{2I},
\end{equation}
where $I$ is the moment of inertia. More generally, we shall see that in an ($N+1$)-dimensional space, the $N-1$ gapped Goldstones with fixed mass gap $\mu$ reproduce the eigenvalues of the Laplacian operator on the $N$-sphere $S^{N}$, \textit{i.e.}~$J(J+N-1)$. A similar check was carried out in Ref.~\cite{Monin:2016jmo} in the large charge limit $J \gg 1$. 

Starting with the three-dimensional case, we consider a point particle parametrized by three coordinates $\vec{q}$, with canonical kinetic energy and a quartic potential expressed in the convenient form
\begin{equation}
V(q)= \dfrac{m^{2}}{2}\left(\left\vert \vec{q} \,\right\vert^{2} - \ell^{2} \right)^{2}.
\end{equation}
In the limit $m\rightarrow \infty $, this corresponds to a quantum mechanical particle confined on the $2$-sphere: $\left\vert \vec{q} \,\right\vert^{2} = \ell^{2}$.
The Hamiltonian is symmetric under ${\rm SO}(3)$ rotations and the associated current is $J^{a}= p_{i}\epsilon^{a}_{\;ij} q_{j}$. Choosing a fixed direction $\bar{a}=3$, we add a chemical potential term for $J^{\bar{a}}$ to obtain the generalized Hamiltonian $H_{\mu} = H -\mu J^{3}$. After straightforward computations the finite $\mu$ Lagrangian can be expressed as:
\begin{equation}
L_{\mu}= \dfrac{1}{2} \dot{q}_{i}\dot{q}^{i} + \mu \epsilon^{3}_{\;ij} \dot{q}_{i} q_{j}-  \dfrac{m^{2}}{2}\left(\left\vert \vec{q} \,\right\vert^{2} - \ell^{2} \right)^{2} + \dfrac{1}{2} \mu^{2} \left( q_{1}^{2}+ q_{2}^{2} \right) .
\end{equation}
For positive $m^{2}$ the ground state of the classical potential $V(q)$ is obtained for\footnote{Without loss of generality, we choose the ground state to be aligned along the $q_{1}$ direction.}
\begin{equation}
q_{1,\rm min}= \ell \dfrac{m^{2}}{m^{2}-\mu^{2}} \xrightarrow[m \rightarrow +\infty]{} \; q_{1,\rm min} = \ell, \qquad  q_{2,\rm min} = q_{3,\rm min}=0 .
\end{equation}
The corresponding moment of inertia is given by $I= \vec{q}_{\rm min}^{\,2}= \ell^{2}$, where we used that the classical Lagrangian $L_{\mu}$ describes the motion of a point particle with unit mass. The average angular momentum and the corresponding ground state energy (expressed as a function of $J$ after a Legendre transform) are:
\begin{equation}
\begin{split}
&J_{m\rightarrow\infty}^{\rm tree}= - \dfrac{\partial}{\partial \mu} V_{\rm tree}(\vec{q}_{\rm min};\mu) =\mu \ell^{2} , \\
&E_{0}^{\rm tree}(J)= J\mu(J) + V_{\rm tree}(\vec{q}_{\rm min};\mu(J)) = \dfrac{J^{2}}{2 \ell^{2}}  +{\rm const},
\end{split}
\end{equation}
so that up to a zero-point energy renormalization, the leading order relation $E_{0}=J^{2}/2I$ is recovered. 
Moreover, we see that the large charge limit corresponds to the limit $\mu \ell^{2} \rightarrow \infty$, where we are varying $\mu$ while keeping $\ell$ fixed.

The subleading result is reproduced by the one-loop computation. The analysis of section~\ref{sec:qm} goes through with minor modifications and the one-loop contribution is given by the frequencies $\omega_{i}$, the poles of the propagator of the quadratic action obtained by expanding around the tree level minimum $q_{\rm min}$. We find:
\begin{equation}
\omega_{1}=0, \qquad \qquad \omega_{2} = \mu, \qquad \qquad \omega_{3} = m .
\end{equation} 
The first two poles correspond to the gapless and the gapped Goldstone excitations~\cite{Nicolis:2012vf}, whereas the third one is the massive radial mode, as expected on general grounds.

The value of the one-loop effective potential at the minimum is
\begin{equation}
V_{\rm eff}(\vec{q}_{\rm min};\mu) = V_{\rm tree}(\vec{q}_{\rm min};\mu)  + \dfrac{1}{2} \sum_{i} \omega_{i} \xrightarrow[m \rightarrow +\infty]{} \; - \dfrac{1}{2} \ell^{2} \mu^{2} + \dfrac{m}{2} + \dfrac{\mu}{2} , 
\end{equation}
which implies
\begin{equation}
\begin{split}
&J_{m\rightarrow\infty}= - \dfrac{\partial}{\partial \mu} V_{\rm eff}(\vec{q}_{\rm min};\mu) =\mu \ell^{2} - \dfrac{1}{2} , \\
&E_{0}(J)= J\mu(J) + V_{\rm eff}(\vec{q}_{\rm min};\mu(J)) = \dfrac{J(J+1)}{2 \ell^{2}}  +{\rm const},
\end{split}
\label{rr1}
\end{equation}
where as before $\ell^{2}=I$.
This reproduces the well-known quantization condition for the eigenvalues of a quantum mechanical rigid rotor, and connects our approach to the (dual) large charge approach of~\cite{Monin:2016jmo}. Notice that in the approach of the present paper, with fixed chemical potential, the correct quantization condition is obtained without taking the large charge limit.

More generally, if we consider a rigid rotor confined on an $N$-sphere $S^{N}$ and introduce a chemical potential for one component of the current associated to the ${\rm SO}(N+1)$ symmetry, from the counting of Goldstone bosons of Ref.~\cite{Nicolis:2012vf} we can immediately infer that there will be one gapless Goldstone, $N-1$ gapped Goldstones with fixed gap $\mu$, plus the massive radial mode. Repeating the previous analysis we obtain
\begin{equation}
E_{0}(J)= \dfrac{J(J+N-1)}{2 I}  +{\rm const},
\label{rr2}
\end{equation}
reproducing the eigenvalues of the Laplacian of the $N$-sphere.

\section{Alternative derivation of the path integral for $\mu<m$}
\label{app:muindependence}

We provide here an alternative derivation of the $\mu$ independence of the path integral for $\mu<m$. We need to compute the integral in eq.~\eqref{eq:iplus}.
Expanding the $\log$ as a sum of two terms we have:
\begin{equation}
I_{+}=\int \dfrac{{\rm d}^{d}p}{(2\pi)^{d}} \left[ \log \left(p^{2} +m^{2}+\xi_{0}^{2}\right) + \log \left(1+ \dfrac{2 \xi_{0} p_{0}}{p^{2} +m^{2}+\xi_{0}^{2}}\right) \right] .
\end{equation}
We can Taylor expand the second term\footnote{The inequality $(p_{0}-\xi_{0})^{2} \geq 0 \Rightarrow p_{0}^{2}+\xi_{0}^{2}\geq 2\xi_{0} p_{0}$ guarantees that we are inside the radius of convergence of the expansion.} to obtain:
\begin{equation}
I_{+}=\int \dfrac{{\rm d}^{d}p}{(2\pi)^{d}}  \log \left(p^{2} +m^{2}+\xi_{0}^{2}\right) + \int \dfrac{{\rm d}^{d}p}{(2\pi)^{d}} \sum_{n=1}^{\infty} \dfrac{(-1)^{n+1}}{n}  \left(\dfrac{2 \xi_{0} p_{0}}{p^{2} +m^{2}+\xi_{0}^{2}}\right)^{n}.
\end{equation}
Exchanging the integral and the series, and transforming the $(p_{0})^{n}$ integral into a spherically symmetric integral by introducing a compensating factor (see Appendix \ref{app:integrals}):
\begin{align}
I_{+}&=\int \dfrac{{\rm d}^{d}p}{(2\pi)^{d}}  \log \left(p^{2} +m^{2}+\xi_{0}^{2}\right) + \sum_{n=1}^{\infty} \dfrac{(-1)^{n+1}}{n} (2 \xi_{0} )^{n} \int \dfrac{{\rm d}^{d}p}{(2\pi)^{d}}   \left(\dfrac{p_{0}}{p^{2} +m^{2}+\xi_{0}^{2}}\right)^{n} \nonumber\\
&=\int \dfrac{{\rm d}^{d}p}{(2\pi)^{d}}  \log \left(p^{2} +m^{2}+\xi_{0}^{2}\right) - \sum_{k=1}^{\infty} \dfrac{1}{2 k} (2 \xi_{0} )^{2k} \dfrac{\Gamma(\frac{d}{2}) \Gamma(k+\frac{1}{2})}{\Gamma(\frac{1}{2})\Gamma(k+\frac{d}{2})}\int \dfrac{{\rm d}^{d}p}{(2\pi)^{d}} \dfrac{p^{2k}}{(p^{2} +m^{2}+\xi_{0}^{2})^{2k}}\nonumber\\
&=  -\dfrac{\Gamma(2-\frac{d}{2})}{\frac{d}{2}(\frac{d}{2}-1)} \dfrac{1}{(4\pi)^{d/2}} \Lambda^{4-d} (m^{2}+\xi_{0}^{2})^{\frac{d}{2}} 
\nonumber\\ &\qquad\qquad
 - \sum_{k=1}^{\infty} \dfrac{1}{2 k} (2 \xi_{0} )^{2k} \dfrac{\pi^{d/2}}{(2\pi)^{d}} \dfrac{\Gamma(k+1/2) \Gamma(k-d/2)}{\Gamma(1/2) \Gamma(2k)} \dfrac{\Lambda^{4-d}}{(m^{2}+\xi_{0}^{2})^{k-d/2}}.
\end{align}
Resumming and taking the limit $d\rightarrow 4-2 \varepsilon$ we obtain
\begin{equation}
I_{+}=\dfrac{m^4 \left(\log \left(\frac{m^4}{\Lambda^{4}}\right)+2 \gamma -3- \log
   (16\pi^{2} )\right)}{64 \pi ^2}-\frac{m^4}{32 \pi ^2 \epsilon },
\end{equation}
in agreement with the result obtained previously.

\section{One-loop path integral at finite $\mu$ for arbitrary potential}
\label{app:det}

In this appendix we derive a general expression for the determinant relevant for the computation of the one-loop effective potential of a complex scalar field at finite $\mu$ with arbitrary interactions in $d$ dimensions.

We use a notation in terms of real components for the scalar fields $\varphi_1,\varphi_2$ and consider the finite $\mu$ Lagrangian derived in section~\ref{sec:scalar_mu}:
\begin{equation}
\mathcal{L}_{\mu}=  \dfrac{1}{2} (\partial_{\nu}\varphi_{1})^{2} +\dfrac{1}{2} (\partial_{\nu}\varphi_{2})^{2} + \mu \left( \dot{\varphi_{1}}\varphi_{2}- \dot{\varphi_{2}}\varphi_{1}\right) - V(\varphi;\mu),
\end{equation}
with an arbitrary ${\rm U}(1)$ invariant interaction potential
\begin{equation}\label{eq:tree_mu_potential}
V(\varphi;\mu) = \dfrac{m^2-\mu^2}{2} (\varphi_{1}^2+\varphi_{2}^2)+ V_{\rm int}(\varphi).
\end{equation}
We shall denote partial derivatives $\partial_{\varphi_i}$ of the function $V(\varphi;\mu)$ with a subscript $i$ and suppress the argument $\mu$ for ease of notation.
Up to boundary terms, the quadratic action for the perturbations $\delta \varphi_{i}(x)$ around an arbitrary constant and homogeneous field background $\varphi_{i}$ is:
\begin{equation}
\delta S^{(2)}[\delta \varphi_{i}(x)]=-\dfrac{1}{2} \int {\rm d}^d x \, 
\delta \vec{\varphi}^{\,t} \, \mathbf{K} \,
 \delta \vec{\varphi},
\end{equation}
where in position space
\begin{equation}
\mathbf{K}(x)=\begin{pmatrix}\vspace{0.5em}
\partial_{\nu}\partial^\nu + V_{11}(\varphi) & -2\mu \partial_{t} +V_{12}(\varphi) \\[0.5em] 
 2\mu \partial_{t} + V_{12}(\varphi) & \partial_{\nu}\partial^\nu + V_{22}(\varphi)
 \end{pmatrix},
\end{equation}
while in momentum space
\begin{equation}
\mathbf{K}(p)=\begin{pmatrix}\vspace{0.5em}
-p^2 + V_{11}(\varphi) & 2i \mu \, p_0 + V_{12}(\varphi) \\[0.5em] 
- 2i \mu \,p_0  + V_{12}(\varphi) & -p^2 + V_{22}(\varphi)
 \end{pmatrix}.
\end{equation}
We can now compute the one-loop effective action through the path integral as
\begin{equation}
e^{i \, \Gamma(\varphi)} = e^{- i \, {\rm Vol} \cdot V_{\rm eff}(\varphi)} =\underset{\delta q_{i}(\pm \infty)=0} {\int}{\hspace{-15pt}[\mathcal{D}\delta \varphi_{i}(x)}] e^{i \,\delta S^{(2)}[\delta \varphi_{i}(x)]} = \left[\det \left(\dfrac{i}{\pi} \mathbf{K} \right)\right]^{-1/2}.
\end{equation}
As discussed in section~\ref{sec:iepsilon}, the $i\varepsilon$ term projects on the ground state of $V(\varphi;\mu)$, and the structure of the poles is such that the integration contours can be deformed continuously, without crossing any singularity, by performing a Wick rotation and going to Euclidean space. We can therefore set  $p_{E}^{0}= -ip^0$ and $p_{E}^{2}= - p^2$. Dropping the $E$ subscript and defining $\xi = ( i \mu, \vec{0})$ to maintain a formally Lorentz invariant notation, and up to an additive constant, the one-loop effective potential is given by
\begin{equation}
\label{eq:veff}
V_{\rm eff}^{(1)}(\varphi;\mu) = \dfrac{1}{2} \int \dfrac{{\rm{d}}^{d}p}{(2\pi)^{d}} \log \left[(p^{2}+M^{2})^{2} - 4 (p \cdot \xi)^{2} - g^2 \right],
\end{equation}
where we defined
\begin{equation}
\begin{split}
&M^{2}=\dfrac{1}{2}\Big(V_{11}(\varphi) + V_{22}(\varphi)\Big) , \\
&g^2= \dfrac{1}{4}\Big(V_{11}(\varphi) + V_{22}(\varphi)\Big)^2 - \Big(V_{11}(\varphi)\Big) \Big(V_{22}(\varphi)\Big)+ \Big(V_{12}(\varphi)\Big)^2. 
\end{split}
\end{equation}
It is straightforward to check that the same result is recovered when using a notation in terms of the complex scalar $\Phi$.

It is useful to express these quantities in terms of the interaction potential:
\begin{equation}\label{eq:Mg1}
\begin{split}
&M^{2}=m^2-\mu^2 +\dfrac{1}{2}\Big(\partial_{\varphi_1}^2 V_{\rm int}(\varphi) + \partial_{\varphi_2}^2  V_{\rm int}(\varphi)\Big) , \\
&g^2= \dfrac{1}{4}\Big(\partial_{\varphi_1}^2  V_{\rm int}(\varphi) + \partial_{\varphi_2}^2V_{\rm int}(\varphi)\Big)^2 - \Big(\partial_{\varphi_1}^2  V_{\rm int}(\varphi)\Big) \Big(\partial_{\varphi_2}^2 V_{\rm int}(\varphi)\Big)+ \Big(\partial_{\varphi_1}\partial_{\varphi_2} V_{\rm int}(\varphi)\Big)^2, 
\end{split}
\end{equation}
from which it follows that the combination $(M^2+\mu^2)$ is $\mu$ independent, and $g^2$ is $\mu$ independent and vanishes for $\varphi_i=0$.

Another, particularly useful, representation is obtained by expressing $M^2$ and $g^2$ in terms of $\phi =[(\varphi_{1}^{2}+\varphi_{2}^{2})/2]^{1/2}$. After straightforward manipulation we find:
\begin{equation}\label{eq:Mg2}
\begin{split}
&M^{2}=\dfrac{1}{4\,\phi}\Big(V'(\phi) + \phi \,V''(\phi)\Big) , \\
&g^2= \dfrac{1}{16\,\phi^2}\Big(V'(\phi) - \phi \,V''(\phi)\Big)^2,
\end{split}
\end{equation}
where primes denote derivatives with respect to $\phi$. Denoting by $\phi_{0}$ the value of $\phi$ at which the tree level finite $\mu$ potential~\eqref{eq:tree_mu_potential} is minimized --- such that $V'(\phi_{0})=0$ ---  it follows that:
\begin{equation}
M^2\Big\vert_{\rm min} = g\Big\vert_{\rm min} = \dfrac{V''(\phi_{0})}{4}.
\end{equation}
This fact proves to be extremely useful in finding a closed form expression for the one-loop low-energy effective action for the superfluid phase of our UV scalar field theory, which can be obtained by evaluating the effective potential for the UV theory at finite $\mu$ at its minimum.

\section{Some useful identities in dimensional regularization}
\label{app:integrals}

The following general identities on dimensionally regularized integrals turn out to be very useful. For a thorough introduction to the methods and properties of dimensional regularization see for instance~\cite{Collins:1984xc}.

\begin{align}
\int \dfrac{{\rm{d}}^{d}p}{(2\pi)^{d}} \log \left(p^{2}+\Delta \right) & = - \dfrac{\Gamma(-d/2)}{(4\pi)^{d/2}} \Delta^{d/2} \\ \nonumber \\
\int \dfrac{{\rm{d}}^{d}p}{(2\pi)^{d}} \dfrac{p^{\mu_1}p^{\nu_1}\dots p^{\mu_n}p^{\nu_n}}{\left(p^{2}+\Delta \right)^m} & = \dfrac{\Gamma(d/2)\Gamma(n+1/2)}{\Gamma(1/2)\Gamma(n+d/2)} \, \eta^{(\mu_1\nu_1}\dots \eta^{\mu_n\nu_n)}  \int \dfrac{{\rm{d}}^{d}p}{(2\pi)^{d}} \dfrac{p^{2n}}{\left(p^{2}+\Delta \right)^m} \\ \nonumber \\
 \int \dfrac{{\rm{d}}^{d}p}{(2\pi)^{d}} \dfrac{p^{2n}}{\left(p^{2}+\Delta \right)^m} & = \dfrac{1}{(4\pi)^{d/2}} \dfrac{\Gamma(m-n-d/2)\Gamma(n+d/2)}{\Gamma(m)\Gamma(d/2)} \left(\dfrac{1}{\Delta}\right)^{m-n-d/2} 
\end{align}

\section{Divergencies and counterterms at finite $\mu$}
\label{app:divergencies}

As we discussed in section~\ref{sec:gexpansion} the divergent terms arising in the one-loop effective potential in $d=4$ are $\mu$ independent. In particular, defining 
\begin{equation}
\Xi^{(1)}(\phi;\mu)\Big\vert_{d,{\rm div}} \equiv V_{\rm eff}^{(1)}(\phi;\mu)\Big\vert_{d,{\rm div}} - V_{\rm eff}^{(1)}(\phi;0)\Big\vert_{d,{\rm div}},
\end{equation}
one finds that $\Xi^{(1)}=0 $ in $d=4$. Starting from $d=6$, on the other hand, the divergencies turn out to be $\mu$ dependent, see Table~\ref{tab:divergencies}. 

\begin{table}[h!]
\centering
\begin{tabular}{c|c|c}
 & $\varepsilon \cdot V_{\rm eff}^{(1)}(\phi;\mu)\Big\vert_{d,{\rm div}}$ & $\varepsilon \cdot \Xi^{(1)}(\phi;\mu)\Big\vert_{d,{\rm div}}$ \\[2ex]
\hline
\hline
\rule{0pt}{1cm} $d=4+\varepsilon$ & $\dfrac{g^2 + (M^2 + \mu^2)^2}{16 \pi^2}$ &  0  \\[3ex]
\hline
\rule{0pt}{1cm} $d=6+\varepsilon$  & $- \dfrac{(M^2+\mu^2)^3+g^2(3M^2+\mu^2)}{192\pi^3}$ &  $\dfrac{g^2}{96\pi^3}\mu^2$    \\[3ex]        
\hline
\rule{0pt}{1cm} $d=8+\varepsilon$  & $\dfrac{5g^4 + 5 (M^2+\mu^2)^4+g^2(30M^4+20 M^2\mu^2 +6 \mu^4)}{15360\pi^4}$ &  $- \dfrac{g^2(M^2+\mu^2)}{384\pi^4}\mu^2+ \dfrac{g^2}{960\pi^4}\mu^4$  \\[3ex]        
\end{tabular}
\caption{\it Divergent terms in dimensional regularization in even dimensions $d=2n+\varepsilon$ (with $n>1$), and their $\mu$ dependent part. See eqs.~\eqref{eq:Mg1} and~\eqref{eq:Mg2} for the expression of $M^2$ and $g^2$ in terms of the interaction potential $V_{\rm int}(\phi)$ and their $\mu$ dependence.}
\label{tab:divergencies}
\vspace{0.5cm}
\end{table}
\begin{figure}[h!]
\begin{center}
\includegraphics[width=6in]{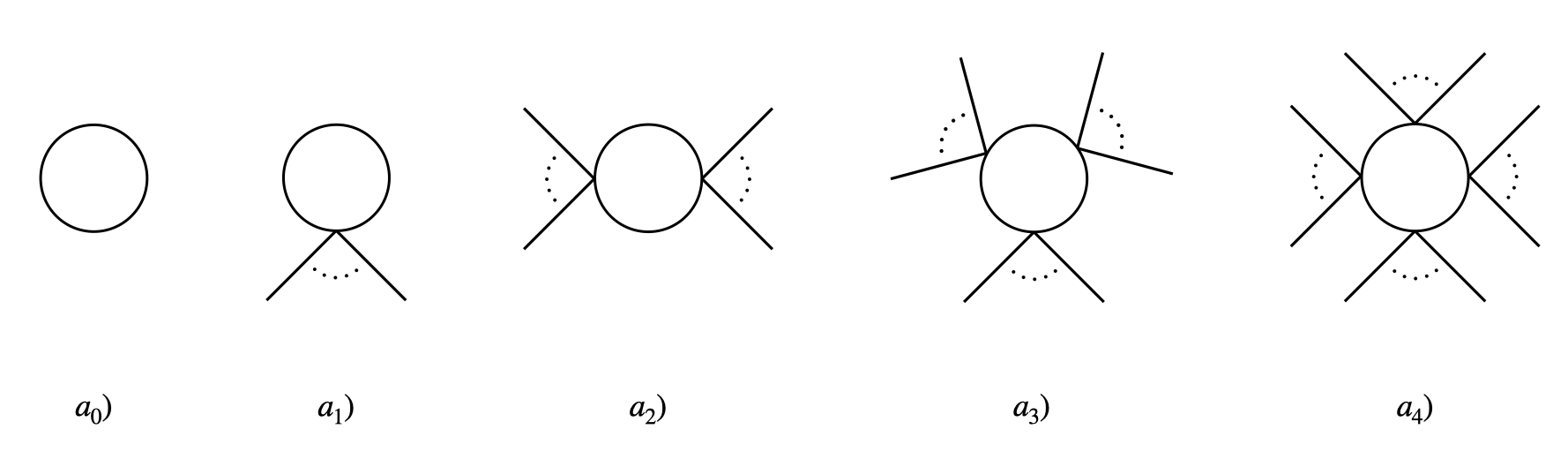}
\end{center}
\caption{\label{fig:diagrams} \it One loop Feynman diagrams up to four interaction points. Diagrams $a_0)$ and $a_1)$ are divergent in dimension $d>2$, but the corresponding divergence is always independent of external momenta. Diagram $a_n)$ is marginally (logarithmically) divergent in $d=2n$ dimensions. Derivative counterterms in the $\mu=0$ theory arise from diagrams $a_n)$ with $n\geq 2$ when they are at least quadratically divergent. This corresponds to $d\geq6$.}
\end{figure}

In order to understand this, consider the $\mu=0$ theory. We take $V_{\rm int}(\phi)$ to be an arbitrary ${\rm U}(1)$ invariant polynomial potential, built from even powers of $\phi$. Since we are considering (possibly) non-renormalizable interactions, the counterterms needed to renormalize the theory will in general include derivative operators, even if the interaction terms we started from are non-derivative. Introducing a chemical potential is equivalent to promoting ordinary derivatives to $\mu$ dependent covariant derivatives, so that derivative counterterms will generate $\mu$ dependent counterterms as well.\footnote{We are grateful to Riccardo Rattazzi for a discussion on this point.} By a simple power counting argument it is easy to see that no derivative counterterms are needed at one loop in $d=4$, for arbitrary interactions. In order to see this notice that the (amputated) diagrams which depend on external momenta, $a_n)$ with $n\geq 2$ in Fig.~\ref{fig:diagrams}, are convergent or at most logarithmically divergent in $d=4$. As a consequence, the corresponding divergencies are independent of external momenta and can be renormalized by non-derivative polynomial counterterms.
This is a generalization of the familiar statement that at one loop there is no wave-function renormalization in $d=4$. In the case of even ($\rm U(1)$ symmetric) interactions the wave-function renormalization is still absent at one-loop for $d=2n$, but starting from $d=6$ derivative interaction counterterms are needed. Notice that interaction terms are always irrelevant in $d \geq 6$, so that the theories we consider are non-renormalizable and the derivative operators we are discussing are always generated.

In order to check our diagrammatic argument, consider the example of the $\lambda \phi^4$ theory in $d=6$. The $\mu$ dependent divergence in the one-loop effective potential can be read from Tab.~\ref{tab:divergencies} and is given by 
\begin{equation}\label{eq:mudivd6}
\Xi^{(1)}(\phi;\mu)\Big\vert_{d,{\rm div}} = \dfrac{1}{\varepsilon} \dfrac{\lambda^2}{24\pi^3} \mu^2 \phi^4.
\end{equation}
On the other hand, in the $\mu=0$ theory, the only diagram that can have divergencies dependent on external momenta is $a_2)$ in Fig.~\ref{fig:diagrams}. Computing the divergent part of this diagram we find that the following derivative counterterm is needed for $\mu=0$:
\begin{equation}
\mathcal{L}_{\rm ct} \supset - \dfrac{1}{\varepsilon} \dfrac{\lambda^2}{24\pi^3}  (\partial_{\nu}\Phi)^{\dagger}(\partial^{\nu}\Phi) \Phi^{\dagger}\Phi,
\end{equation}
where we only wrote the divergent part. Promoting derivatives to $\mu $ dependent covariant derivatives $D_\nu= \partial_{\nu} -i \mu  \delta_\nu^0$, we find that a $\mu$ dependent counterterm is generated that exactly cancels the divergence of eq.~\eqref{eq:mudivd6}.

As a last remark, we note that the classic argument on the non-renormalization of conserved currents (see \emph{e.g.}~\cite{Collins:1984xc}, p. 162) advocated in~\cite{Benson:1991nj} to justify the absence of $\mu$ dependent divergencies in $d=4$ is not applicable at finite $\mu$. In fact, at finite chemical potential, an identically conserved current can be constructed with the aid of the object $\mu  \delta_\nu^0$, invalidating the non-renormalization theorem. The current is simply given by $\tilde{J}^{\nu}~=~(\partial^{\nu} \partial_0 - \delta^{\nu}_{0}\square) \phi^2$.

\section{More details on the derivation of the scaling relations of section~\ref{sec:scaling_relations}}
\label{app:details}
In this appendix the variable $X$ will be always set at its background value $X\rightarrow \bar X=\mu^2$.
We start from the relations $f^2_\pi(\Q) = \Q/\mu = 2P'(X)$ and $\Q= 2P'(X) \sqrt{X}$, and the sound speed
\begin{equation}
c_s^2= \dfrac{P'(X)}{P'(X)+2P''(X)X}.
\end{equation}
The derivatives of $f_\pi^2$ are:
\begin{align}
&\dfrac{{\rm d}f_\pi^2}{{\rm d}\Q} = \dfrac{{\rm d} (2P'(X))}{{\rm d}X} \dfrac{{\rm d}X}{{\rm d}\Q} = 2 P''(X) \dfrac{1}{2P''(X)\sqrt{X} + P'(X)/\sqrt{X}}= \dfrac{1-c_s^2}{\sqrt{X}},\\
&\dfrac{{\rm d^2}f_\pi^2}{{\rm d}\Q^2} = - \dfrac{{\rm d}c_s^2}{{\rm d}\Q} \dfrac{1}{\sqrt{X}} + (1-c_s^2) \dfrac{{\rm d} }{{\rm d}X}\left(\dfrac{1}{\sqrt{X}}\right) \dfrac{{\rm d} X}{{\rm d}\Q} = - \dfrac{{\rm d}c_s^2}{{\rm d}\Q} \dfrac{1}{\sqrt{X}} - \dfrac{c_s^2}{\Q} \dfrac{ (1-c_s^2) }{\sqrt{X}}.
\end{align}
The derivative of the sound speed is instead:
\begin{equation}
\dfrac{{\rm d}c_s^2}{{\rm d}\Q} = \dfrac{{\rm d}c_s^2}{{\rm d}X} \dfrac{{\rm d}X}{{\rm d}\Q} = \dfrac{{\rm d}c_s^2}{{\rm d}X} \dfrac{1}{2P''(X)\sqrt{X} + P'(X)/\sqrt{X}}.
\end{equation}
In the limit $\Q\rightarrow 0$, using that $2P'(X)=\mathcal{O}(\Q)$, we obtain:
\begin{equation}
\dfrac{{\rm d}c_s^2}{{\rm d}X} =  \dfrac{P''(X)}{P'(X)+2P''(X)X} + \mathcal{O}(\Q) = \dfrac{1}{2X}  + \mathcal{O}(\Q) \implies \dfrac{{\rm d}c_s^2}{{\rm d}\Q} = \dfrac{1}{4 X^{3/2} P''(X)} + \mathcal{O}(\Q) .
\end{equation} 
Using that for $\Q\rightarrow 0$, $\sqrt{X}=\mu_{\rm crit} +  \mathcal{O}(\Q)  = m_{\rm pole} +  \mathcal{O}(\Q) $ we arrive at the result:
\begin{equation}
\dfrac{{\rm d}c_s^2}{{\rm d}\Q} = \dfrac{1}{4 m_{\rm pole}^{3/2} P''(m_{\rm pole}^2)} + \mathcal{O}(\Q) .
\end{equation}
Higher order terms can be straightforwardly computed by rewriting $c_s^2$ as
\begin{equation}
c_s^2= \dfrac{\Q}{\Q+4P''(\mu^2)\mu^3},
\end{equation}
and then expressing $\mu(\Q)$ recursively through 
\begin{equation}
\mu(\Q)=  m_{\rm pole} \exp\left(\int_0^Q\frac{c_s^2}{\Q'}{\rm d}Q'\right).
\end{equation}

\section{Ground state and stability for the ${\rm O}(N)$ model with quartic interactions }
\label{app:SON}

We want to study the minimum of the large $N$ effective potential in the ${\rm O}(N)$ model with quartic interactions in $d=3$, given by eq.~\eqref{eq:eff_potential_SON}. The conditions for minimization are given in eqs.~\eqref{eq:conditions}. First, it is easy to see that the minimum of the effective potential always corresponds to $S_i=0$. Moreover it is convenient to define the $\rm SO(2)$ invariant variable $t= (s_1^2+s_2^2)/N$. The stationary points of $V_{\rm eff}(s_i)$ as a function of $s_i$ correspond through the chain rule to either $t=0$ or to real stationary points of $V_{\rm eff}(t)$, satisfying $V'_{\rm eff}(t)=0$. As already remarked in~\cite{Coleman:1974jh}, it can be convenient to solve the condition~\eqref{eq:condition1} for $\chi$ and plug back in eq.~\eqref{eq:eff_potential_SON} to study the ordinary effective potential, without the auxiliary parameter $\chi$. From eq.~\eqref{eq:condition1} we have:
\begin{equation}
f(t) \equiv \sqrt{\chi} = \dfrac{\lambda_N}{8\pi} \left(\sqrt{1+ \dfrac{64\pi^2}{\lambda_N}\left(t+\dfrac{m^2}{\lambda_N}\right)} - 1 \right),
\end{equation}
where we selected the solution with $\sqrt{\chi}>0$. We obtain
\begin{equation}
\dfrac{V_{\rm eff}(t)}{N} = \dfrac{1}{2} \left(f^2 -\mu^2 \right) t - \dfrac{\left(f^2 - m^2 \right) ^2}{4\lambda_N} - \dfrac{f^3}{12\pi},
\end{equation}
from which it follows 
\begin{equation}\label{eq:VprimeSON}
\dfrac{V'_{\rm eff}(t)}{N} = \dfrac{1}{2} \left(f^2 -\mu^2 \right) + ff' t- \dfrac{\left(f^2 - m^2 \right)}{\lambda_N} f f' - \dfrac{f^2 f'}{4\pi} =  \dfrac{1}{2} \left(f^2 -\mu^2 \right) ,
\end{equation}
where we suppressed the explicit $t$ dependence in $f(t)$ and used $f^2-m^2 = \lambda_N t - \lambda_N f/4\pi $. The stationary points of $V_{\rm eff}(t)$ correspond to real (hence physical) values of $t$ only when 
\begin{equation}
\mu^2 \geq \mu_{\rm crit }^2 = f(0)^2,
\end{equation}
from which eq.~\eqref{eq:mucrit_largeN} follows. For values of $\mu^2 \leq \mu_{\rm crit }^2$ the potential $V_{\rm eff}(s_i)$ is minimized for $t=0$, corresponding to the unbroken phase. Moreover we see that in $d=3$ and for $\mu^2 > \mu_{\rm crit }^2$ the potential has only one stationary point for $t > 0$ (a minimum) and is always bounded from below, thus stable. The situation is different in $d=4$, where due to logarithmic terms the potential is unbounded from below, signalling an instability~\cite{Gross:1974jv,Coleman:1974jh}.

Notice also that by rescaling eq.~\eqref{eq:VprimeSON} we can also derive the value of the pole mass in the unbroken phase of the UV theory at $\mu=0$. This is given by 
\begin{equation}
m^2_{\rm pole} = \dfrac{2}{N} V'_{\rm eff}(t;\mu=0) = f(0)^2 = \left( \sqrt{\dfrac{\lambda_N^2}{64\pi^2}+m^2} - \dfrac{\lambda_N}{8\pi}\right)^2.
\end{equation}
In particular, it follows that $\mu_{\rm crit}^2=m_{\rm pole}^2$. This result is non-perturbative in $\lambda_N$, at leading order in the $1/N$ expansion.

\phantomsection
\addcontentsline{toc}{section}{References}
\bibliographystyle{utphys}
{\linespread{1.075}
\bibliography{biblio}
}

\end{document}